\documentclass[preprint]{vldb}
\pdfoutput=1
\usepackage[activate={true,nocompatibility}]{microtype}
\usepackage[all]{nowidow}

\usepackage{etex}
\usepackage{balance}
\usepackage{booktabs} % For formal tables
\usepackage[usenames, dvipsnames]{color}
\usepackage[table]{xcolor}
\usepackage{subcaption}
\usepackage{hyperref}
\usepackage{listings}
\usepackage{multirow}
\usepackage{enumitem}
\usepackage{textcomp}
\usepackage[T1]{fontenc}
\usepackage{upquote}
\usepackage{balance}
\usepackage[font=small]{caption}
\usepackage{tcolorbox}
\usepackage{soul}
\usepackage{times}
\usepackage{xspace}
\usepackage[sort,nocompress]{cite}
\usepackage{amsthm}
\usepackage{pifont}% http://ctan.org/pkg/pifont
\usepackage{tikz}
\usepackage{etextools}
\usepackage[normalem]{ulem}
\usetikzlibrary{arrows}

\makeatletter
\def\@copyrightspace{\relax}
\def\@copyrightblurb{\relax}
\def\@mkbibcitation{\relax}
\makeatother

\makeatletter
\def\thm@space@setup{\thm@preskip=2pt
\thm@postskip=2pt}
\makeatother

\SetTracking[ spacing = { , , } ]{ encoding = T1, family = tt* }{ -50 }
\newcommand{\code}[1]{\ifuppercase{#1}{\scalebox{0.95}{\textls{\texttt{#1}}}}{\texttt{#1}}}

\newcommand{\stitle}[1]{\vspace{0.25em}\noindent\textbf{#1}}

\newcommand{\eat}[1]{}

\newcommand{\modin}[0]{{\sc Mo\-din}\xspace}
\newcommand{\stmt}[0]{statement\xspace}
\newcommand{\stmts}[0]{statements\xspace}

\newcommand{\techreport}[1]{#1}

\newcommand{\papertext}[1]{}
\newcommand{\rebuttal}[1]{#1}
\newcommand{\removed}[1]{} % Using st isnt very stable for multiple lines in the code.
\newcommand{\moved}[1]{#1}

\newcommand{\squishlist}{
   \begin{list}{$\bullet$}
    { \setlength{\itemsep}{0pt}
      \setlength{\parsep}{2pt}
      \setlength{\topsep}{2pt}
      \setlength{\partopsep}{0pt}
      \setlength{\leftmargin}{12pt}
    }
}
\newcommand{\squishend}{\end{list}}

\definecolor{newc}{RGB}{70, 180, 80}
\definecolor{pinegreen}{rgb}{0.0, 0.47, 0.44}

\newcommand{\jeg}[1]{\ignorespaces}
\newcommand{\agp}[1]{\ignorespaces}
\newcommand{\devin}[1]{\ignorespaces}
\newcommand{\william}[1]{\ignorespaces}
\newcommand{\simon}[1]{\ignorespaces}
\newcommand{\dorx}[1]{\ignorespaces}
\newcommand{\dlee}[1]{\ignorespaces}
\newcommand{\adj}[1]{\ignorespaces}
\newcommand{\omkar}[1]{\ignorespaces}
\newcommand{\rehan}[1]{\ignorespaces}
\newcommand{\jmh}[1]{\ignorespaces}
\newcommand{\smacke}[1]{\ignorespaces}

\newcommand{\todo}[1]{\ignorespaces}

\def\pandas{pandas\xspace}
\def\Pandas{Pandas\xspace}

\vldbTitle{}
\vldbAuthors{Authors}
\vldbDOI{https://doi.org/10.14778/3352063.3352148}
\vldbVolume{12}
\vldbNumber{12}
\vldbYear{2019}

\begin{document}
% \input{modin-rebuttal}
%\vldbPages{XXXX-YYYY}

\title{Towards Scalable Dataframe Systems\papertext{\\
{\Large [Vision Paper]}}
}
 
%%%%%%%%%      Authors       %%%%%%%%%%%%%%%%%%%%
\numberofauthors{1}

\author{
\alignauthor
 Devin Petersohn, 
        Stephen Macke, Doris Xin,
        William Ma, Doris Lee, 
        Xiangxi Mo \\ Joseph E.~Gonzalez,
        Joseph M.~Hellerstein,
        Anthony D.~Joseph, 
        Aditya Parameswaran \\
    UC Berkeley \\
    \email{\tt \{devin.petersohn, smacke, dorx, williamma, dorislee, xmo, jegonzal, hellerstein, adj,  adityagp\}
    @berkeley.edu}
}

\maketitle
%\sccaps{%

\begin{abstract}
\rebuttal{Dataframes are a popular abstraction to represent, prepare, and analyze data.
Despite the remarkable success of dataframe libraries in R and Python, dataframes face performance issues even 
on moderately large datasets. 
Moreover, there is significant ambiguity regarding dataframe semantics.
In this paper we lay out a vision and roadmap for scalable dataframe systems.
To demonstrate the potential in this area, we report
on our experience building \modin, 
a scaled-up implementation of the most widely-used and complex 
dataframe API today, Python's \texttt{pandas}.
With pandas as a reference, we propose a simple
data model and algebra for dataframes to ground discussion in the field. 
Given this %common
foundation, 
we lay out an agenda of open research opportunities where
the distinct features of dataframes will require
%adapting and
extending the state of the art in many dimensions of data 
management.
% This agenda includes new problems in data storage, metadata management, query 
% optimization and processing, and interactive data processing.
% This agenda includes new problems in data storage, metadata management, query 
% optimization and processing, and interactive data processing 
We discuss the implications of signature dataframe features
including flexible schemas, ordering, row/column equivalence, and 
data/metadata fluidity, as well as the piecemeal, trial-and-error-based
approach to interacting with dataframes. 
}

 \end{abstract}
 % Section 1
%!TEX root = modin-vision.tex

\section{Introduction}\label{sec:intro}

For all of their commercial successes, 
relational databases have notable limitations 
when it comes to 
``quick-and-dirty'' exploratory data analysis 
(EDA)~\cite{tukey1977exploratory}. Data needs to be 
defined schema-first before it can be 
examined, 
data that is not well-structured is 
difficult to query, 
and any query beyond
\code{SELECT *}
requires an intimate 
familiarity with the schema, which is particularly problematic for wide tables. 
% % Begin additional EDA comments
% Exploratory Data Analysis (EDA) often involves examining and querying un-
% or semi-structured data where types may be dynamically inferred or
% specified, making it difficult and sometimes impossible to define the schema
% upfront.
% % End additional EDA comments
For more complex analyses, the 
declarative nature of SQL makes it
awkward to develop and debug queries in a piecewise, modular
fashion, conflicting with best practices for software development.
\rebuttal{In part thanks to these limitations, SQL is
often not the tool of choice for data exploration. 
As an alternative,}
programming languages such as Python and R support 
the so-called {\em data\-frame}
abstraction. Dataframes provide a functional interface 
that is more tolerant
of unknown data structure and well-suited 
to developer and data scientist workflows, 
including REPL-style imperative interfaces 
and data science notebooks~\cite{perez2015project}.

Dataframes have 
several characteristics 
that make them an appealing 
choice for data exploration:

\squishlist
\item an
intuitive data model that 
embraces an implicit ordering on both rows
and columns and treats them symmetrically;
\item a query 
language that bridges a variety of data 
analysis modalities including 
relational (e.g., filter, join), 
linear algebra (e.g., transpose), 
and spreadsheet-like (e.g., pivot) operators; 
\item an incrementally composable query 
syntax that encourages easy and rapid validation
of simple expressions, 
and their iterative refinement and composition
into complex queries; and 
\item native embedding in a 
host language such as Python 
with familiar imperative semantics.
\squishend

\noindent 
% These characteristics and others have helped
% dataframes become the de-facto
% standard for EDA. \jmh{citation? can we find one
% that says they're more popular than SQL in this regard} 
% In particular, the dataframe abstraction provided
% by 
Characteristics such as these
have helped dataframes become
incredibly popular for EDA; 
for instance, 
the dataframe abstraction provided by \pandas 
within Python (\url{pandas.pydata.org}), 
has, as of 2020, been downloaded
over 300 million times, served as a dependency for
over 222,000 repositories in GitHub, 
and starred on GitHub more 25,000 times.
Python's own popularity
has been 
attributed to the success of \pandas 
for data exploration and data science~\cite{python-growing-quickly,python-growth-blamed}.
Due to its ubiquity, we focus on \pandas for concreteness.

\Pandas 
has been developed from the ground up
\rebuttal{via open-source contributions
from dozens of contributors,
each providing operators and
their implementations to the DataFrame API}
to satisfy immediate or ad-hoc needs, 
spanning capabilities
that mimic
relational algebra, 
linear algebra, and 
spreadsheet computation.
To date, 
the \pandas DataFrame API has ballooned to 
over 200 operators~\cite{pandas-api}.
R, which is 
both more mature and more carefully curated, has only 70 operators---but this
still far more than, say, relational
and linear algebra combined~\cite{R-dataframe-api}.

While this rich API is sometimes cited as a 
reason for \pandas' attractiveness,
the set of operators has significant redundancies,
often with different performance implications.
These redundancies place a considerable burden 
on users to select the optimal way of expressing their goal.
For example, one blog post cites five different ways to express
the same goal, with performance varying from 0.3ms to 600ms 
(a 1700$\times$ increase)~\cite{optimizing-pandas};
meanwhile, the \pandas documentation itself offers multiple recommendations
for how to enhance performance~\cite{enhancing-performance}.
As a result, many users eschew the bulk of the API, 
relying only on a small
subset of operators~\cite{minimally-sufficient}. 
The complexity of the API and evaluation semantics
also make it difficult to apply 
traditional query optimization techniques.
Indeed, each operator within a \pandas ``query plan'' 
is executed completely before subsequent operators are executed,
with limited optimization, and 
no reordering of operators or pipelining 
(unless explicitly done so by the user using \code{.pipe}). 
Moreover, the performance of the \code{pandas.DataFrame} API
breaks down when processing 
even moderate volumes
of data that do not fit in memory, as we will see 
subsequently---this is especially problematic
due to \pandas' eager evaluation semantics, wherein
intermediate data items often surpass main memory limits
and must be paged to disk.

\rebuttal{
To address \pandas' scalability challenges,
we developed \modin (\url{github.com/modin-project/modin}), 
our first attempt at a scalable dataframe system,
which employs 
parallel query execution
%\jmh{rather than ``simple optimizaton'' say ``parallel query execution''?}
to 
enable unmodified \pandas code 
to run more efficiently on large dataframes.
\modin is used by over 60 downstream projects, and
has over 250 forks and 4,800 stars on GitHub\techreport{ in its first 20 months}, indicating
the impact and need for such systems.
\modin rewrites \pandas API calls 
into a sequence
of operators in a new, compact 
dataframe algebra.
\modin then leverages simple 
parallelization and a new physical 
representation 
to speed up the 
execution of these operators,
by up to {\bf 30$\times$} in certain
cases, and is able to run to completion
on datasets {\bf 25$\times$} larger 
than \pandas %(which crashes instead) 
in others.
}

\rebuttal{
Our initial optimizations in \modin are
promising, but only scratch the surface of what's possible. Given that 
first experience and the popularity of the results, 
we believe
there is room for a {\bf \em broad, community
research agenda on
% However, {\bf \em \modin only represents
% our first step towards 
% a broad research agenda for
making dataframe systems
scalable and efficient}, with many novel research challenges. 
Our original intent when developing \modin
was to adapt standard relational database techniques
to help make dataframes scalable.
However, while the principles (such as parallelism)
do apply, their instantiation in the form of specific 
techniques often differ,
thanks to the differences between
the data models and algebra of dataframes and relations.
Therefore, a more principled
foundation for dataframes is needed,
comprising a formal data model
and an expressive and compact algebra.
We describe our first attempt
at such a formalization in Section~\ref{sec:definitions}.
Then, armed with our data model
and algebra,
we outline a number of 
research challenges organized around 
unique dataframe characteristics and
the unique ways in which they are processed.

In Section~\ref{sec:datamodel},
we describe how the dataframe data model and algebra
result in new scalability challenges.
Unlike relations, dataframes
have a flexible schema and are lazily typed,
requiring careful maintenance of metadata,
and avoidance of the overhead of type inference as
far as possible.
Dataframes treat rows and columns
as equivalent, and metadata (column/row labels) and data 
as equivalent, requiring flexible ways
to keep track of metadata and orientation,
placing new metadata awareness
requirements on dataframe query planners to avoid physically transposing data where possible.
In addition, dataframes are ordered---and dataframe systems
often enforce a strict coupling between logical and physical
layout; we identify several opportunities to deal
with order in a more light-weight, decoupled, and lazy fashion. 
Finally, the new space of operators---encompassing
relational, linear algebra, and spreadsheet operators---introduce
new challenges in query processing and optimization.

In Section~\ref{sec:interactivity},
we describe new challenges and opportunities
that emerge from how dataframes are used
for data exploration.
Unlike SQL which 
offers an all-or-nothing query modality, 
dataframe
queries are 
constructed one operator
at a time, with ample think-time between query fragments.
This makes it more challenging to perform query optimization
wherein operators can be reordered for higher overall efficiency.
At the same time, the additional thinking time between steps
can be exploited to do background processing. 
Users often inspect intermediate dataframe results 
of query fragments, usually
for debugging, which requires a costly materialization
after each step of query processing.
However, users are only shown an ordered 
prefix or suffix of this intermediate
dataframe as output, allowing
us to prioritize the execution to return
this portion quickly and defer the execution
of the rest. 
Finally, users often revisit old processing steps in 
an ad-hoc process of trial-and-error data exploration.
We can consider opportunities to minimize redundant computation
for operations completed previously.

% Inspired by our experiences
% with \modin, we propose a new
% research agenda centered around the development
% of scalable dataframe systems.
% \modin simply represents our first step
% for this agenda---many interesting outstanding
% research challenges remain in making
% dataframe systems scalable and efficient. 
\stitle{Outline and Contributions.}
In this paper, 
we begin with an example dataframe workflow
capturing typical 
dataframe capabilities and user behaviors.
We then 
describe our experiences with \modin 
(Section~\ref{sec:modin}).
We use \modin to ground our discussion of
the research challenges.
We {\bf \em (i)
% define a research agenda around 
% formalizing
provide a candidate formalism for
dataframes} and enumerate their capabilities with a new algebra
(Section~\ref{sec:definitions}).
We then outline research challenges and 
opportunities
 {\bf \em to build on our formalism
and make dataframe systems more scalable,}
by optimizing 
and accounting
for 
{\bf \em (ii) the unique characteristics 
of the new data model and algebra} (Section~\ref{sec:datamodel}),
as well as {\bf \em (iii)
the unique ways in which dataframes are used 
in practice for data exploration} (Section~\ref{sec:interactivity}).
We draw on tools
and techniques from the database research literature throughout
and discuss how they might be adapted 
to meet novel dataframe needs.
}

In describing the aforementioned challenges, we focus on the \pandas dataframe system~\cite{pandas-api} 
for concreteness. 
\Pandas is much more popular than other dataframe 
implementations,
and is therefore well worth our effort to study and optimize.
We discuss other dataframe
implementations and related work in Section~\ref{sec:ext}.
\papertext{Many details about \modin and our dataframe data model and algebra are 
omitted 
and can be found in our technical report~\cite{petersohn2020towards}.}

\section{Dataframe Example}\label{sec:usecase}

\begin{figure*}[!t]
 \vspace{-10pt}
    \centering
    \includegraphics[width=0.98\textwidth]{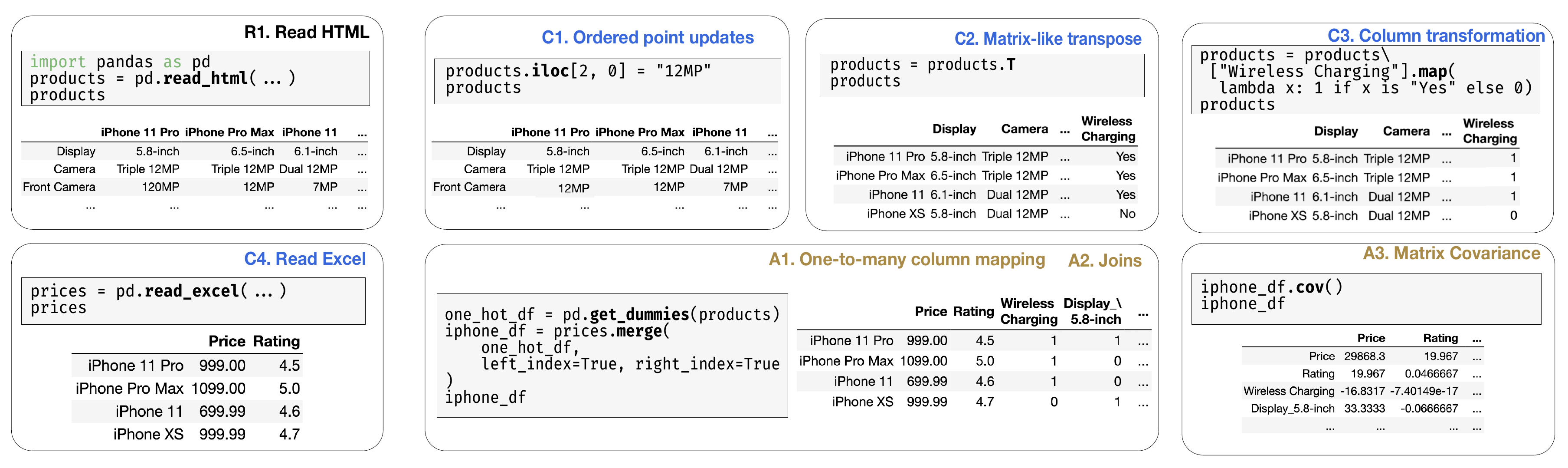}
     \vspace{-10pt}
    \caption{\rebuttal{Example of an end-to-end data science workflow, from data ingestion, preparation, wrangling, to analysis. }
    %The variable names of the dataframes being printed are indicated on the left, and the Cleaning or Analysis phase (denoted by C or A, respectively) are indicated on the right.
    % \agp{\tiny We need a third column for the output post the read html. read\_html entirety needs to be in bold. we can depict the change for C1 via a strikethrough with a new value overlaid on the old df. likewise for C3 (strikethroughs for every value in the column) overlaid on the old df. anyone volunteering? A1/A2: what happened to the phone type column? A3 needs ... for the missing columns. }
    }
    \label{fig:example}
     \vspace{-15pt}
\end{figure*}

% \removed{
% Consider an analyst
% exploring the relationship 
% between price, rating, and product
% features across different iPhone models. 
% Figure~\ref{fig:example}
% illustrates the steps taken and the results in the context
% of a Jupyter Notebook.
% % william: uncomment when resolve rebuttal (\url{www.jupyter.org})
% With this example,
% we aim to illustrate key aspects 
% of a typical dataframe workflow.
% %  and contrast
% % it with relational query processing.
% }

\rebuttal{
In Figure~\ref{fig:example},
we show the steps taken
in a typical workflow of an analyst exploring the relationship between
various features of different iPhone models in a Jupyter notebook~\cite{perez2015project}.
}

\stitle{Data ingest and cleaning.} 
Initially, the analyst reads in the
iPhone comparison chart \rebuttal{using \code{read\_html}}
from an e-commerce webpage, as shown in 
R1 in Figure~\ref{fig:example}. 
% \agp{Call this R1 in figure and refer to it as such}
% \removed{
% The analyst prints out
% the first few lines of the dataframe 
% to verify that the \pandas data
% loading function call 
% resulted in the expected dataframe.
% They do so by simply calling \code{products}. 
% We omit this call 
% and all subsequent calls to display
% dataframes to save space,
% and instead depict the call 
% alongside the corresponding 
% output.
% }
\rebuttal{
The data is verified
by printing out the first few lines of the dataframe
\code{products}. (\code{products.head()} is also often used.)
}
Based on
% \rebuttal{Using} % William: Noun-verb disagreement.
this preview of the dataframe,
the analyst identifies a sequence of
actions for cleaning their dataset:
\squishlist
    \item C1 [Ordered point updates]: 
        % \removed{
        % The front camera description for the iPhone 11 Pro 
        % has an anomalous value of 120MP 
        % (indicated in Figure~\ref{fig:example} in {\bf bold} for emphasis), 
        % far exceeding that of other models---likely a data entry error. 
        % To fix this, 
        % the analyst performs spreadsheet-like point updates, 
        % by setting a new value for the entry at row 2 column 0 via an \code{iloc} command, 
        % as in \code{In[2]}, and then verifies the output in \code{Out[2]}.
        % }
        \rebuttal{
        The analyst fixes the anomalous value of 120MP for Front Camera
        for the iPhone 11 Pro to 12MP, 
        by performing a point update via \code{iloc}, and views the result. 
        }
    \item C2 [Matrix-like transpose]: 
        % \removed{
        % To make it easier to compare across different products, 
        % the rows and columns should be switched, 
        % so that the rows correspond to products and 
        % the attributes correspond to features.
        % The analyst makes this switch by transposing the data
        % via \code{.T} in \code{In[3]} and then inspects the results in \code{Out[3]}.
        % }
        \rebuttal{
        To convert the data to a relational format, 
        rather than one meant for human consumption,
        the analyst transposes the dataframe (via \code{T}) 
        so that the rows are now products and
        columns features,
        and then inspects the output.
        }
    \item C3 [Column transformation]: 
        % \removed{
        % %The string column "Wireless Charging" should be converted from Yes/No to 1/0.
        % Since the dataset was scraped from a webpage intended for human consumption,
        % the column labeled ``Wireless Charging'' contains ``Yes/No'' 
        % instead of binary values, 
        % the latter of which may be preferable for downstream processing.
        % The analyst makes this change 
        % by setting an existing column label to the
        % result of a user defined \code{map} function
        % that appropriately modifies the data.
        % Using \code{map} to update columns is common 
        % during wrangling and cleaning, as depicted in \code{In[4]},
        % with the output dataframe inspected in \code{Out[4]}.
        % }
        \rebuttal{
        The analyst further modifies the dataframe 
        to better accommodate downstream data processing by
        changing the column ``Wireless Charging'' from ``Yes/No'' to binary.
        This is done by updating the column using a user-defined \code{map} function, followed by displaying the output.
        }
    \item C4 [Read Excel]: 
        % \removed{
        % To integrate a different dataset into the workflow, 
        % the analyst now loads in an additional dataset from a spreadsheet
        % containing price and rating information for various iPhone models,
        % depicted in \code{In[5]}, 
        % and inspects the resulting dataframe \code{prices} in \code{Out[5]}.
        % }
        \rebuttal{
        The analyst loads price/rating information
        by reading it from a spreadsheet into \code{prices} and then
        examines it.
        }
\squishend

\stitle{Analysis.} 
\rebuttal{Then,} the analyst performs the following operations to analyze the data:
% \agp{joe says A1 is still prep; will need to fix fig}

\squishlist
    \item A1 [One-to-many column mapping]: 
        % \removed{
        % The analyst encodes categorical features as binary indicators(0/1) 
        % across multiple columns via the one-hot encoding scheme. 
        % As shown in line 1 of \code{In[6]}, 
        % this encoding is invoked via the \code{get\_dummies} function,
        % with the results in \code{one\_hot\_df}.  
        % }
        \rebuttal{
        The analyst encodes non-numeric features in a one-hot encoding scheme
        via the \code{get\_dummies} function.
        }
    \item A2 [Joins]: 
        % \removed{
        % To combine the information in \code{prices}
        % with the product features in \code{one\_hot\_df}, 
        % the analyst performs a left inner join via \code{.merge()} 
        % on the column corresponding to iPhone model names in line 2 of \code{In[6]},
        % and inspects the results in \code{Out[6]}.
        % }
        \rebuttal{
        The iPhone features are joined
        with their corresponding price and rating
        using the \code{merge} function.
        The analyst then verifies the output.
        }
    \item A3 [Matrix Covariance]: 
        With all the relevant numerical data in the same dataframe,
        the analyst 
        %\remove{now} 
        computes the covariance between the features
        via 
        %\remove{a call to \code{df.cov(...)}}%
        \rebuttal{the \code{cov} function, and examines the output}.
\squishend

\noindent This example demonstrated 
only a sample of the capabilities of dataframes.
Nevertheless, it serves to illustrate 
the common use cases for dataframes:
immediate visual inspection after most operations, 
each incrementally building on the results of previous ones,
point and batch updates
via user-defined functions,
and a diverse set of operators for wrangling,
\rebuttal{preparing,} and analyzing data. 
% Moreover, dataframes offer additional flexibility
% for data manipulation compared to traditional RDBMSs.
% Next, we formally define the dataframe
% data model, and query language.
% Section 3
%!TEX root = modin-vision.tex

\section{The Modin Dataframe System}\label{sec:modin}

% \removed{
% \modin 
% % \footnote{\modin's name comes from the Korean
% % word for every, as in every \pandas operator or every
% % dataframe API.}
% is our open-source  
% dataframe system that implements 
% the data model and algebra described 
% in Section~\ref{sec:definitions}. 
% }

\rebuttal{
While the \pandas API is convenient and powerful,
the underlying implementation has many scalability
and performance problems.
We therefore started an effort to develop
a ``drop-in'' replacement for the \pandas API, 
\modin\footnote{\scriptsize \modin's name is derived from the Korean
word for ``every'', as it targets every dataframe operator.},
to address these issues. 
In the style of embedded database systems~\cite{sqlite2020hipp,raasveldt2019duckdb},
Modin is a library that runs in the same process as the application that
imports it.
We briefly describe the challenges we encountered
and the lessons we learned
during our implementation in Section~\ref{sec:optim},
followed by a preliminary case of \modin's performance in Section~\ref{sec:accelerate}. 
% We then discuss takeaways from our experience
% developing \modin and how that informs
% our future research agenda in Section~\ref{subsec:takeaways}.
\papertext{We defer detailed treatment of \modin's architecture
to our technical report~\cite{petersohn2020towards}.}
\techreport{Finally, we describe \modin's architecture
and implementation.}
}

\subsection{\rebuttal{Modin Engineering Challenges}}\label{sec:optim}
\rebuttal{
% We now discuss the engineering
% challenges we encountered when trying 
% to make \pandas more scalable.
When we started our effort to make \pandas
more scalable,
we identified that while many operations in \pandas
are fast, they are limited by their
single-threaded implementation.
Therefore, our starting point for \modin 
was to add multi-core capabilities and other
simple performance improvements to enable 
\pandas users to run their same unmodified workflows
both faster and on larger datasets.
However, we encountered a number of engineering
challenges.
}

\stitle{Massive API.}
\rebuttal{The \pandas API 
has over 240 distinct operators,
making it challenging to individually optimize
each one. 
After manually trying to parallelize
each operator within \modin,
we tried a different approach. 
We realized that there is a lot of redundancy
across these 240 operators.
Most of these operators can be rewritten
into an expression composed using 
a much smaller set of operators.
We describe our compact set of dataframe
operators---our working dataframe algebra---in 
Section~\ref{sec:df-algebra}.
Currently, \modin supports over 85\% of
the \code{pandas.DataFrame} API,
by rewriting API calls into our working algebra,
allowing us to avoid duplicating 
optimization logic as much as possible.
The operators we prioritized
were based on an analysis of
over 1M Jupyter notebooks\papertext{, the results of
which are discussed in our technical report~\cite{petersohn2020towards}.}
\techreport{discussed in Section\ref{sec:usage}.}
Specifically, we targeted all the functionality in
\code{pandas.DataFrame}, \code{pandas.Series}, and
\pandas  utilities (e.g., \code{pd.\-concat}).
To use \modin instead of \pandas, users can simply 
invoke ``\code{import modin.pandas}'',
instead of ``\code{import pandas}'', and proceed
as they would previously.
}
\techreport{
\modin is implemented in Python using over 30,000 lines
of code. 
\modin is completely open source and can be found at 
\url{https://github.com/modin-project/modin}.}

\stitle{Parallel execution.}
\rebuttal{Since most \pandas
operators are single-threaded,
we looked towards parallelism
as a means to speed up execution.}
Parallelization
is commonly used to improve performance 
in a relational context 
due to the embarrassingly parallel nature 
of relational operators.
Dataframes have 
% \removed{some unique properties and requirements}
\rebuttal{a different set of operators
than relational tables, supporting relational algebra,
linear algebra, and spreadsheet operators,
as we saw in Section~\ref{sec:usecase}, and we
will discuss in Section~\ref{sec:definitions}.
We implemented different internal mechanisms for 
exploiting parallelism} 
depending on the 
data dimensions and operations being performed. 
\rebuttal{Some} operations
are embarrassingly parallel and
can be performed on each row independently \rebuttal{(e.g., C3 in Figure~\ref{fig:example})},
\rebuttal{while others (e.g., C2, A1, A3) cannot.}
% \removed{Our current simple approach to partitioning is to 
% do it on a per-operation basis, and to pick between}
\rebuttal{To address the challenge of differing levels
of parallelism across operations, we designed
\modin to be able to flexibly
move between common partitioning schemes:}
row-based (i.e., each partition has a collection of rows), 
column-based (i.e., each partition has a collection of columns), 
or block-based partitioning 
(i.e., each partition has a subset of rows and columns),
depending on the operation.
\rebuttal{Each partition is then processed independently
by the execution engine, with the results communicated
across partitions as needed.}

\begin{figure*}[ht!]
    \centering
    \vspace{-10pt}
    \includegraphics[width=0.9\textwidth]{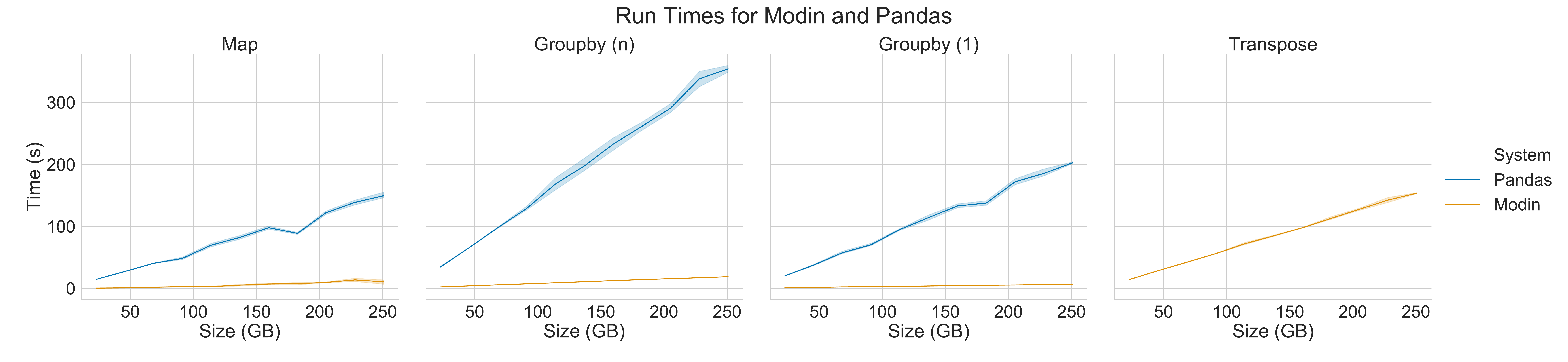}
    \vspace{-10pt}
    \caption{%
        \rebuttal{For each function, 
        we show the runtime for both \modin and \pandas 
        and the 95\% confidence interval.
        There are no times for transpose with \pandas as 
        \pandas is unable to run transpose beyond 6 GB.}
    }
\label{fig:performance}
\vspace{-15pt}

\end{figure*} 
 
\stitle{Supporting billions of columns.}
\rebuttal{While parallelism does address
some of the scalability challenges, it fails
to address a major one: the ability to support
tables with billions of columns---something
even traditional database systems do not support.
Using the \pandas API, however, it is possible
to transpose a dataframe (as in Step C2)
with billions of rows 
into one with billions of columns.}
In many settings, e.g., when dealing with graph
adjacency matrices in neuroscience or genomics,
the number of rows and number of columns can both be very large.
For these reasons, \modin treats
rows and columns essentially equivalently,
\rebuttal{a property of dataframes will discuss in detail in Section~\ref{sec:definitions}}.
In particular, to transpose a large dataframe,
\modin employs block-based partitioning,
where each block consists of a 
subset of rows and columns.
Each of the blocks are individually transposed, 
followed by a simple change of the overall metadata 
tracking the new locations of each of the blocks.
The result is a transposed dataframe that does not
require any communication.

\subsection{Preliminary Case Study}\label{sec:accelerate}

\rebuttal{
To understand how the simple optimizations discussed above
impact the scalability of dataframe operators,
we perform a small case study evaluating}
\modin's performance against 
that of \pandas using 
microbenchmarks on an EC2 x1.32xlarge (128 cores and 1,952 GB RAM) 
node using a New York City taxicab dataset~\cite{nyc-taxi-data} that was replicated 1 to 11 
times to yield a dataset size 
between 20 to 250 GB, with up to 1.6 billion rows.
We consider four queries: 
\squishlist
\item map: check if each value in the dataframe is null, and replace it with a \code{TRUE} if so, and \code{FALSE} if not. 
\item groupby ($n$): 
%\code{GROUP BY(DF, "passenger\_count", lambda d: d.count())} - 
group by the non-null ``passenger\_count'' column and count the number of 
rows in each group.
\item groupby ($1$): 
%\code{GROUP BY(DF, 1, lambda d: d.count())} -
count the number of non-null rows in the dataframe.
\item transpose: 
%\code{MAP(TRANSPOSE(DF), lambda d: d)} - 
swap the columns and rows of the dataframe and 
apply a simple (map) function across
the new rows. 
\squishend

% \eat{\agp{rewrite these into the algebra}
% \code{df.groupby().count()}, \code{df.count()}, and \code{df.trans\-pose().map(lambda s: s)},
%  representing the performance of the operators
% \code{GROUP BY}, \code{WINDOW}, and \code{TRANSPOSE} + \code{MAP}, respectively.}
\noindent
We highlight the difference 
between group by with one group and 
$n$ groups, because 
with $n$ groups 
\rebuttal{
data shuffling and
communication are a
factor in performance.
With groupby(1),
the communication overheads
across groups are non-existent.
}
We include transpose 
to demonstrate that \modin
can handle data with 
billions of columns.
\techreport{This query also 
shows where \pandas crashed or 
did not complete in more than 2 hours.}

Figure~\ref{fig:performance} shows that 
for the group by ($n$) 
and group by ($1$) operations,
\modin yields a speedup of up to $19\times$ 
and $30\times$ relative to \pandas, respectively. 
For example, a group by ($n$) on a 250GB dataframe, 
\pandas takes about 359 seconds 
and \modin takes 18.5 seconds, a speedup of more
than 19$\times$.
% \agp{jmh asks: on 128 cores?}
For map operations, \modin is about 12$\times$
faster than \pandas.
These performance gains come from
simple parallelization of operations within \modin,
while \pandas only uses a single core.
During the evaluation of transpose, 
\pandas was unable to transpose 
even the smallest dataframe of 20 GB ($\sim$150 million rows)
after 2 hours.
Through separate testing, 
we observed that \pandas can only transpose 
dataframes of up to 6 GB ($\sim$6 million rows) on the
hardware we used for testing.

\rebuttal{\stitle{Takeaways.} Our preliminary case study and our experience
with \modin
demonstrates the promise of integrating
simple optimizations to make dataframe systems scalable.
Next, we define a dataframe data model
and algebra to allow us to ground our subsequent discussion of
our research agenda, targeting the unique characteristics
of dataframes and the unique ways in which they are used.
We defer further performance analyses of \modin to  future work.}

\techreport{
\subsection{The MODIN Architecture}\label{sec:architecture}

\begin{figure}
\centering
\vspace{-5pt}
\includegraphics[width=.4\textwidth]{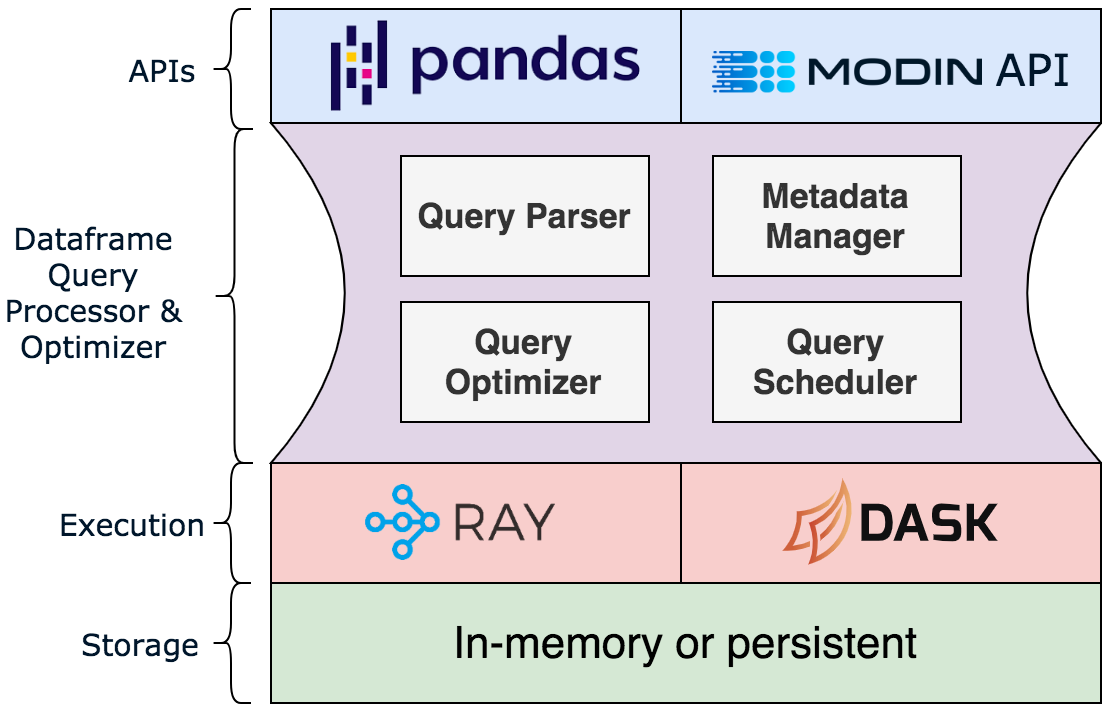}
\vspace{-5pt}
\caption{\modin architecture.}
\label{fig:modin-arch}
\vspace{-15pt}
\end{figure}

\modin's architecture is {\em modular} 
for easy integration of new storage and execution engines, APIs, and optimizations. It consists of four layers: 
the API layer, the query processing
and optimization layer, 
the execution layer, 
and the storage layer, shown in Figure~\ref{fig:modin-arch}.

\stitle{API layer.} 
Users can leverage
\modin via a \pandas-based API,
or directly via a leaner and simpler \modin API
based on the algebra in Section~\ref{sec:df-algebra}.
In either case, the API layer
translates each call into 
a dataframe algebraic expression,
and passes that to the next layer for execution.
The layer isolates users from changes
to the layers below, while allowing
users to leverage the API modality they are most
comfortable with. 
\techreport{Future implementations may support other user
APIs for working with dataframes,
such as SQL or relational algebra.}
Our \pandas-based API currently supports
about 150 of over 200 \pandas dataframe APIs,
and rewrites each of them into dataframe
algebraic expressions.

\stitle{Query processing and optimization layer.}
As shown in Figure~\ref{fig:modin-arch}, 
the query processing layer follows a
``narrow waist'' design, 
exposing a small API based on 
the dataframe algebra,
and implements the data model 
from Section~\ref{sec:dataframedef}.
This layer parses, optimizes, and executes 
dataframe queries with the help
of layers below. 
As we will describe in Section~\ref{sec:optim},
\modin leverages parallel execution 
of dataframe queries on multiple
dataframe partitions, 
scheduled on execution engines
in the next layer. 
This layer also keeps track of 
dataframe metadata  
including row labels, column labels, 
and column data types.
\techreport{Recall that data types may not be specified
on dataframe creation, so }
\modin
induces types on-the-fly (using the $S$ function)
when needed for a specific operation.

\stitle{Execution layer.} \modin
supports distributed 
processing of 
dataframe partitions
using 
two execution frameworks: 
Ray~\cite{ray} and Dask~\cite{dask}.
Both Ray and Dask are task-parallel asynchronous 
execution engines exposing an API that requires defining a task or function and providing data for the task to run on. 
Integration of a new
execution framework is simple, often requiring
fewer than 400 lines of code.

\stitle{Storage layer.} 
\modin's modular storage layer
supports both main memory and 
persistent storage out-of-core (also called memory spillover),
allowing intermediate dataframes to exceed main-memory limitations
while not throwing memory errors, unlike \pandas. 
To maintain \pandas semantics, the dataframe
partitions are freed from persistent storage once
a session ends.
}

% Section 4
%!TEX root = modin-vision.tex

\section{Dataframe Fundamentals}\label{sec:definitions}
% \smacke{Maybe call this ``Dataframe Background''?
% Or simply ``Dataframes''?
% }
%  \jmh{I wish there was history here. Is R responsible for the idea? Matlab?
% I have assumed that R is the reference, and
% Pandas is just a terrible implementation
% that we're now unwinding and expanding. 
% I realize we may 
% not want to say so in this paper, but
% knowing/sharing the history would be cool.}\dlee{I'm adding some history related to dataframes into S7.}

There are many competing open-source and commercial
implementations of dataframes, but
there is no formal definition or 
enumeration of dataframe properties 
in the literature to date.
\rebuttal{
We therefore 
propose a formal definition of dataframes
to allow us to describe 
our subsequent research challenges
on a firm footing, and also 
to provide background to readers who are 
unfamiliar with dataframes. 
% We describe this as a part of this vision paper to
% establish a common understanding of the semantics
% and formalism given the relative newness of this abstraction
% to the database community.
}
In this section, we start with a brief history (Section~\ref{sec:dataframe-history}), and 
provide a reference data model (Section~\ref{sec:dataframedef})
and algebra (Section~\ref{sec:df-algebra})
to ground discussion. We then
demonstrate the expressiveness of the
algebra via \rebuttal{a case study} 
(Section~\ref{sec:algebra-example})\techreport{ and 
discuss extensions (Section~\ref{sec:dataframe-extensions}).
We finally provide
some quantitative statistics
into dataframe usage in Section~\ref{sec:usage}}.
\papertext{\rebuttal{Our technical report
has additional details about the formalism,
the mapping to the \pandas API, 
other extensions to the data model,
as well as quantitative statistics on dataframe usage~\cite{petersohn2020towards}}.}

% \subsection{Dataframe Formalization}
% % Dataframes exhibit a number of fundamental deviations
% % from traditional relational tables.
% In this section, we present a formal dataframe data model
% and propose a set of dataframe algebra operators that
% capture the APIs and operators in
% traditional dataframe implementations like \pandas and R.

% \smacke{I like the bit about the inherent symmetry,
% but this still feels a bit too vague in my opinion.
% Since dataframes are best understood by the
% functionality they provide, would it make more sense
% to give a formal specification of properties that
% dataframes exhibit? E.g. 1. two-dimensional ordered data, 2. typing and labels / names attached to both rows and columns,
% 3. symmetry between rows / cols (anything that holds for cols holds for rows and vice versa), 4. weakly schema'd, 5. provides API with host interpreted
% language for use in an interactive environment like
% a Jupyter notebook.}
% We define a dataframe as an {\em ordered} tabular relation \dorx{???}
% \smacke{Maybe say: {\em Dataframes} organize data in two dimensions
% and allow for many of the same operations permitted by traditional relational
% algebra, while also enforcing {\em ordering semantics} in both the rows and
% columns and allowing for additional expressivity due to the inherent
% symmetry between rows and columns.}
\subsection{A Brief History of Dataframes}\label{sec:dataframe-history}

The S programming language was developed at Bell Laboratories
in 1976 to support statistical computation.
Dataframes were first
introduced to S in 1990, and presented by
Chambers, Hastie, and Pregibon at the Computational Statistics
conference~\cite{chambers-compstat-1990}.
The authors state: ``We have introduced into S a class of objects
called \code{data.frames}, which can be used if convenient
to organize all of the variables relevant to a particular 
analysis ...''
Chambers and Hastie then extended this paper into a 1992 book~\cite{chambers1992statistical}, which states
``Data frames are more general than matrices in the sense
that matrices in S assume all elements to be of the same mode---all
numeric, all logical, all character string, etc.'' and ``... data frames support matrix-like
computation, with variables as columns and observations as rows,
and, in addition, they allow computations in which
the variables act as separate objects, referred to by name.''

The R programming language, an open-source implementation of
S\techreport{ with some additional innovations}, 
was first released in 1995,
with a stable version released in 2000, and gained
instant adoption among the statistics community. 
%\william{Is this referring to the R dataframe or the language itself?}
Finally, in 2008, Wes McKinney developed \pandas in an effort to bring
dataframe capabilities with R-like semantics to Python, which as we described
in the introduction, is now incredibly popular. 
\techreport{In fact, \pandas is often
cited as the reason for Python's popularity~\cite{python-growing-quickly,python-growth-blamed}, 
now surpassing Java and
C++~\cite{incredible-growth-python}.}
We discuss other dataframe implementations in 
Section~\ref{sec:ext}.

% \newpage
\subsection{Dataframe Data Model}\label{sec:dataframedef}

As Chambers and Hastie themselves state, 
dataframes are not familiar mathematical objects. 
Dataframes are not quite relations, nor are they matrices or tensors. 
In our definitions we borrow 
textbook relational 
terminology 
from Abiteboul, et al.~\cite[Chapter~3]{abiteboul1995foundations} 
and adapt
it to our use.

\theoremstyle{definition}
\newtheorem{definition}{Definition}[section]
% \begin{definition}{Dataframe:}
% A dataframe is an ordered collection of tabular data
% with inherent symmetry between rows and columns. In
% a dataframe, both rows
% and columns are associated with a label, are ordered,
% and have types.
% \end{definition}

The elements in 
the dataframe
come from a 
known set of 
domains $Dom = \{\mbox{\textbf{dom}}_1, \mbox{\textbf{dom}}_2, ...\}$. 
For simplicity, we assume in our discussion that
domains are taken from the set 
$Dom = \{\Sigma^*$, $\mbox{\textbf{int}}$, $\mbox{\textbf{float}}$, $\mbox{\textbf{bool}}$, 
$\mbox{\textbf{category}}\}$, though a few other useful domains
like datetimes are common in practice. 
The domain $\Sigma^*$ is the 
set of finite strings over an 
alphabet $\Sigma$, and serves as a default, 
uninterpreted domain; 
in some dataframe libraries it is called \textbf{Object}.
Each domain contains a 
distinguished $null$ value, sometimes written as \code{NA}.
Each domain $\mbox{\textbf{dom}}_i$ 
also includes a \emph{parsing function} 
$p_i: \Sigma^* \rightarrow \mbox{\textbf{dom}}_i$,
allowing us to interpret the values in dataframe cells as
domain values\techreport{ (including possibly $null$)}.

A key aspect of a dataframe is that the domains of its columns may
be induced from data \textit{post hoc}, rather than 
being declared \textit{a priori} as in the relational model. 
We define a \textbf{schema 
induction function} $S:{\Sigma^*} \rightarrow Dom$ 
that assigns 
an array of $m$ strings to a domain in $Dom$. 
This schema induction function is applied to
a given column and returns
a domain that 
describes this array of strings; 
we will return to this function later.

\begin{figure}
\vspace{-5pt}
    \centering
    \includegraphics[width=0.3\textwidth]{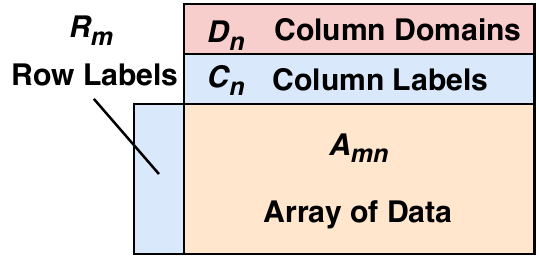}
    \vspace{-5pt}
    \caption{The Dataframe Data Model}
    \label{fig:datamodel}
    \vspace{-15pt}
\end{figure}

Armed with these definitions, we can now define a dataframe:
\begin{definition} % {Dataframe:}
A \textbf{dataframe} is a tuple
$(A_{mn}, R_m, C_n, D_n)$, where $A_{mn}$ is an array of entries 
from the domain $\Sigma^*$,
$R_m$ is a vector of row labels from $\Sigma^*$,
$C_n$ is a vector of column labels from $\Sigma^*$,
and $D_n$ is a vector of $n$ domains from $Dom$, one per column,
each of which can also be left unspecified. 
We call $D_n$ the {\em schema} of the dataframe.
If any of the $n$ entries within $D_n$ 
is left unspecified, 
then that domain can be induced 
by applying $S (\cdot)$ to the corresponding column of $A_{mn}$\techreport{ to get its domain $i$ and then
$p(\cdot)$ to get its values}.
% If $D_n$ is unspecified, it is induced as $[S(A_{mn}^1), \ldots, S(A_{mn}^n)]$, where $A_{mn}^i$ refers to the $i$th column of $A_{mn}$.
\end{definition}
\noindent 
We depict our conceptualization of dataframes
in Figure~\ref{fig:datamodel}.
In our example of Figure~\ref{fig:example}, 
\rebuttal{dataframe \code{products} after step R1} has $R_m$ corresponding to 
an array of labels $[$\code{Display}, \code{Camera}, $\ldots]$;
$C_n$ corresponding to an array of labels
$[$\code{iPhone 11 Pro}, \code{iPhone Pro Max}, $\ldots]$;
$A_{mn}$ corresponding to the matrix of values
beginning with \code{5.8-inch}, with $m = 6, n = 4$.
Here, $D_n$ is left unspecified,
and may be inferred using $S (\cdot)$ per column 
to possibly correspond to $[\Sigma^*, \Sigma^*, \Sigma^*, \Sigma^*]$,
since each of the columns contains strings.
%\agp{What is $D_n$ in this example?}
% \squishlist
% \item Which operators are dependent on $D_n$?
% \item Which operators can be optimized on $D_n$?
% \item Which operators modify $D_n$?
% \item Which operators preserve $D_n$?
% \item $S$ is not user defined
% \item $S$ must touch every row
% \squishend

Rows and columns are symmetric in many ways in dataframes.
Both can be 
referenced explicitly, using either numeric indexing (positional 
notation) or label-based indexing (named notation). 
In our example in Figure~\ref{fig:example}, 
the \code{products} dataframe
is referenced using positional notation in step C1 
with \code{products.iloc[2, 0]} to modify the value in 
the third row and first column, 
and by named notation in step C3 
using \code{products ["Wireless Charging"]} to modify the
column corresponding to \code{"Wireless Charging"}. 
The 
relational model traditionally provides this kind of referencing
only for columns. Note that row position
% ordering 
is exogenous to the data---it \emph{need not} be correlated in
any way to the data values, unlike sort orderings found
in relational extensions like SQL's \code{ORDER BY} clause.
The positional notation allows for $(row, col)$ 
references to index individual values, as is familiar from matrices.

A subtler distinction is that row and column labels are from the 
same set of domains as the underlying data ($Dom$), whereas in the traditional 
relational model, column names are from a separate domain 
(called \textbf{att}~\cite{abiteboul1995foundations}).
This is important to point out because there are dataframe 
operators that
copy data values into labels, or copy labels into data values,
discussed further in Section~\ref{sec:df-algebra}.

\rebuttal{One distinction between rows and columns 
in our model is that columns have a schema, 
but rows do not. 
Said differently, we parse the value of any cell 
based on the domain of its column. 
We can also imagine 
an orthogonal view, in which we define explicit schemas (or use 
a schema induction function) on rows, 
and a corresponding row-wise parsing function 
for the cells. In our formalism,
this is achieved by an algebraic operator to 
transpose the table and treat the result column-wise 
(Section~\ref{sec:df-algebra}). 
By restricting the data model to a single axis of schematization, we 
provide a simple unique interpretation of each cell, 
yet preserve a flexibility of interpretation in the
algebra. In Sections~\ref{sec:matviews} and~\ref{sec:rowcolumneq} 
we return to the performance and programming implications
of programs that make use of schemas on a dataframe and its transpose (i.e. ``both axes'').}
% \rebuttal{One way our data model treats rows and columns as
% different is via $D_n$, which is associated only with columns
% rather than rows. While we could have equivalently
% defined a row-level schema $S_m$, 
% in many cases, e.g., when each of the columns have distinct
% types, this will be overkill, as the typing
% for each row will correspond to an array of different types
% (or more compactly represented as $\Sigma^*$).
% That said, specific dataframe systems may certainly 
% keep track of 
% row schemas, especially when the types of all the values
% in a row is the same, since it will be handy when
% combined with transpose.
% We discuss this choice further in Section~\ref{sec:datamodel}.}

% \removed{
% Despite the notational symmetry between rows and columns,
% $D_n$ introduces a \emph{schematic asymmetry} 
% that we will need to reason about. Consider a schema like
% $D_n = [$\textbf{category}, \textbf{int}, \textbf{float}$]$. Note that
% the column types differ, but the type of each row is the same
% (namely $[$\textbf{category}, \textbf{int}, \textbf{float}$]$). 
% Moreover, each column has a single domain (an atomic type), but 
% each row has a vector of domains (a tuple type). Once seen 
% through the lens of a schema and its parsing functions, rows and columns
% are quite different.
% }

When the schema $D_n$ has the same 
domain \textbf{dom} for all $n$ columns, we call this a \emph{homogeneous} 
dataframe\techreport{, and its
rows and columns can be considered symmetrically to have 
the domain \textbf{dom} differing only in dimension}. 
As a special case, consider a homogeneous dataframe with a domain like 
\textbf{float} or \textbf{int} and operators $+, \times$ that satisfy
the algebraic definition of a field. We call this a \emph{matrix}
dataframe, since it has the algebraic properties required of a matrix,
and can participate in linear algebra operations simply
by parsing its values and ignoring its labels. 
The dataframe \code{iphone\_df} \rebuttal{after step A2}  in Figure~\ref{fig:example}
is one such example; thus
it was possible to perform the covariance operation in step C3.
\rebuttal{
Matrix dataframes are commonly used in machine learning pipelines.
}

\rebuttal{Overall,
while dataframes have roots in both relational and linear algebra,
they are neither tables nor matrices. 
Specifically, when viewed from a relational viewpoint, 
the dataframe data model differs in the following ways:
\begin{table}[!htbp]
\vspace{-5pt}
\small
\centering
\begin{tabular}{ll}
Dataframe Characteristic              & Relational Characteristic                \\ \hline
Ordered table                         & Unordered table                          \\
Named rows labels    & No naming of rows                        \\
A lazily-induced schema               & Rigid schema                             \\
Column names from $d \in Dom$ & Column names from att~\cite{abiteboul1995foundations}                                          \\
Column/row symmetry               & Columns and rows are distinct            \\
Support for linear alg. operators  & No native support
\end{tabular}
\vspace{-5pt}
\end{table}

\noindent And when viewed from a matrix viewpoint,
the dataframe data model differs in the following ways:
\begin{table}[!htbp]
\vspace{-5pt}
\small
\centering
\begin{tabular}{ll}
Dataframe Characteristic                 & Matrix Characteristic                    \\ \hline
Heterogeneously typed                    & Homogeneously typed                      \\
Both numeric and non-numeric types       & Only numeric types                       \\
Explicit row and column labels           & No row or column labels                  \\
Support for rel. algebra operators & No native support
\end{tabular}
\vspace{-5pt}
\end{table}
}

% \todo{Revisit rewriting}
% \noindent \rebuttal{
% We can conceptualize dataframes
% from both relational and linear algebra points of view,
% however the dataframe has some data
% model differences that ultimately conflict the fundamental data model
% of each.

% \stitle{From a relational viewpoint, dataframes consist of the follow data model deviations:}
% \squishlist
%     \item An ordered table
%     \item Named rows of a domain from $Dom$
%     \item A lazily-induced schema
%     \item Column names with a domain from $Dom$
%     \item Column and row symmetry
%     \item Support for linear algebra operators (e.g. matrix multiply)
% \squishend
% }

% \noindent \rebuttal{
% \stitle{From a linear algebra viewpoint, dataframes consist of the follow data model deviations:}
% \squishlist
%     \item Heterogeneous matrix-like data structure
%     \item Both numeric and non-numeric types
%     \item Explicit row and column labels
%     \item Support for relational algebra operators (e.g. join)
% \squishend
% }

% \removed{
% In summary, a dataframe can be viewed in two equivalent ways.
% From a relational viewpoint, dataframes are ordered relations 
% with named rows and a lazily-induced schema.
% From a linear algebra viewpoint, dataframes are heterogeneous matrices
% with added labels for rows and columns.
% }

\noindent 
We will exploit these two viewpoints in our dataframe algebra to
allow us to define both relational and linear algebra operations.
\rebuttal{
Due to these differences, a new body of
work will be needed to support the scale
required for modern data science workflows.}

%  A session is a complete, end-to-end workflow. It 
% begins when the user starts a notebook or interpreter environment
% and ends when the user shuts down that environment. A session
% is composed of one or more queries. Because dataframe systems
% are typically in-memory, the results of the query are not
% persisted unless the user explicitly expresses this by writing
% the result to a file.

\techreport{
\subsubsection*{Data model Comparisons}\label{sec:data-model-comp}
Before we go on, we address some key distinctions 
between dataframes 
and other familiar data models.

\stitle{Comparison with matrices.}
All matrices can be represented as 
dataframes (with null labels).
Not all dataframes can be matrices, however, even if
we strip off their labels! Matrices are
homogeneous in schema, but dataframes allow for schemas
with multiple domains. Even if we ignore the schema, a 
dataframe is still not a matrix---opaque strings from 
$\Sigma^*$ do not satisfy the properties of a field as required 
by a matrix.

\stitle{Comparison with relational tables.}
A relation is defined by a declared schema, and there are 
many possible \emph{instances} of a relation---sets of tuples that 
satisfy the schema. An instance can be thought of as a fixed 
relational table.
Dataframes are something like relation instances: they represent 
a fixed set of data. However their schema can be unspecified
and hence \emph{induced} based on their content by a schema
induction function $S$. This flexibility is critical to dataframes.

Moreover, dataframes impose an ordering and naming on their rows.
Object-oriented relational extensions such as the
Postgres data model also introduced implicit row 
identifiers~\cite{rowe1990postgres}, but typically relational
models do not impose a row ordering.
Of course, we can capture this semantics in a relational model via design discipline: we can ensure that 
all our relations have a unique key (for naming), and
an ordering key (for ordering), which are exogenous to the actual data columns in the table. In this sense, all dataframes
can be represented as relation instances conforming to some 
(potentially induced) schema with appropriate keys.

Even so, a key difference between dataframes
and relations is the symmetry between rows and columns.
This aspect, along with the freedom to induce a schema
on a per-instance basis, 
make it possible to define a transpose operator
on dataframes.
Since the underlying representation
is uninterpreted, we are free to induce a different relational
schema after transposition of a dataframe. 

\stitle{Comparison with spreadsheets.}
At a high level, a spreadsheet is an array
of heterogeneously-typed cells that may contain strings or 
formulae. A spreadsheet thus stores code as well as data. 
The strings are dynamically interpreted into a variety of domains.
It is tempting to think of formula-free spreadsheets as being 
similar to dataframes, given the common row/column indexing 
scheme and the dynamic typing. But in general they are quite different. 
The data representations possible in spreadsheets are quite
free, and in practice often quite irregular.
Dense regions of data are often interspersed with empty regions, 
or with cells containing comments and other forms of human-centric
annotation or metadata.
Spreadsheets
have a notion of a ``range''---a subarray or even a set of subarrays---which
may be sparsely located in the data grid and represent diverse or unrelated data sets stored in the same spreadsheet.
% It is common in the spreadsheet data model to represent multiple
% datasets in the same spreadsheet with some empty cells
% separating them. 
As a result of this freedom of structure, bulk algebraic operations
are difficult to define generally within spreadsheets.
Some modern spreadsheets allow a range to be labeled as a ``table''
that is interpreted with a schema and maintains an ordering;
further extensions to ``pivot tables'' allow for row and column labels. Using these constructs it is possible for a spreadsheet 
to represent one or more dataframes.
But the relative simplicity of dataframes enables a
much simpler algebra and easier
implementation and optimization of the algebra's operators. 
}%techreport

\subsection{Dataframe Algebra} \label{sec:df-algebra}

\begin{table*}[!t]
\centering
\begin{tabular}{|l|c|c|c|c|c|l|}
\hline
\rowcolor[HTML]{C0C0C0} 
\textbf{Operator} & \multicolumn{2}{|c|}{\textbf{(Meta)data}} & \textbf{Schema}    & \textbf{Origin} & \textbf{Order}           & \textbf{Description}                                     \\ \hline
SELECTION         &                   & $\times$        & static & REL             & Parent                   & Eliminate rows                                           \\ \hline
PROJECTION        &                   & $\times$        & static & REL             & Parent                   & Eliminate columns                                        \\ \hline
UNION             &                   & $\times$        & static & REL             & Parent$^\dag$            & Set union of two dataframes                              \\ \hline
%INTERSECTION      &                   & $\times$        & static & REL             & Parent$^\dag$            & Set intersection of two dataframes                       \\ \hline
DIFFERENCE        &                   & $\times$        & static & REL             & Parent$^\dag$            & Set difference of two dataframes                         \\ \hline

CROSS PRODUCT / JOIN             &                   & $\times$        & static & REL             & Parent$^\dag$            & Combine two dataframes by element                        \\ \hline
DROP DUPLICATES   &                   & $\times$        & static & REL             & Parent                   & Remove duplicate rows                                    \\ \hline
GROUPBY          &                   & $\times$        & static & REL             & New            & Group identical \rebuttal{attribute values for a given (set of) attribute(s)}  \\ \hline
SORT              &             &       $\times$        & static  & REL             & New                      & Lexicographically order rows                             \\ \hline
RENAME            & $(\times)         $            &               & static & REL             & Parent                   & Change the name of a column                              \\ \hline
WINDOW            &                   & $\times$        & static & SQL             & Parent                   & Apply a function via a sliding-window (either direction) \\ \hline
TRANSPOSE         & $(\times)$        & $\times$        & dynamic       & DF              & Parent$^\lozenge$        & Swap data and metadata between rows and columns          \\ \hline
MAP               &  $(\times)$                 & $\times$        & dynamic & DF              & Parent                   & Apply a function uniformly to every row                  \\ \hline
TOLABELS         & $(\times)$            & $\times$        & dynamic       & DF              & Parent                   & Set a data column as the row labels column \\ \hline
FROMLABELS      & $(\times)$            & $\times$        & dynamic       & DF              & Parent                   & Convert the row labels column into a data column      \\ \hline
\end{tabular}
\vspace{-5pt}
\caption{
Dataframe Algebra. 
$\dag$: Ordered by left argument first, then right to break ties. 
%$\ddag$: Ordered by parent first, then by the order of the collection. 
$\lozenge$: Order of columns is inherited from order of rows and vice-versa.
}
\label{tab:df-algebra}
\vspace{-10pt}
\end{table*}

\rebuttal{
While developing \modin, we discovered that there exists a
``kernel'' of operators that encompasses the massive
APIs of \pandas and R. We developed this ``kernel'' into
a new dataframe algebra, which we describe here,
while explicitly contrasting it with
relational algebra.
}
We do not
argue that this set of operators is minimal, 
but we do feel it is
both expressive and elegant;
we demonstrate via \rebuttal{a case study} 
in Section~\ref{sec:algebra-example}
\rebuttal{can be used to express \code{pivot};
\techreport{other examples of rewriting for operators
within \pandas can be found in our technical report~\cite{petersohn2020towards}.}}
Based on the contrast with relational algebra, 
we are in a position
to articulate research challenges in optimizing
dataframe algebra expressions in subsequent sections.

\techreport{
To the best of our knowledge,
an algebra for dataframes has never
been defined previously.
Recent work by Hutchinson et al.~\cite{lara, laradb}
proposes an algebra called Lara that combines
linear and relational algebra, exposing 
only three operators:
\code{JOIN}, \code{UNION}, and \code{Ext} (also known as ``flatmap'');
however, the operators below  
that manipulate metadata 
would not be possible in Lara without
placing the metadata as part of the data.
Other differences stem from the flexible data model
and lazily induced schema.
That said, as we continue to refine our algebra,
we will draw on Lara as a reference.
}

We list the algebra operators we have defined in
Table~\ref{tab:df-algebra}: the rows correspond to the operators,
and the columns correspond to their properties.
The operators encompass 
ordered analogs of 
extended relational 
algebra operators (from \code{SELECTION} to \code{RENAME}),
one operator that is not part of 
extended relational algebra but is found in 
many database systems
(\code{WINDOW}), \rebuttal{one operator with that admits independent use
unlike in database systems (\code{GROUPBY}),}
as well as four new operators
(\code{TRANSPOSE}, \code{MAP}, 
\code{TOLABELS}, and \code{FROMLABELS}).
% We briefly describe the first two categories first,
% followed by the \rebuttal{four} new operators.
The ordered analogs of relational algebra 
operators preserve the ordering of the input dataframe(s).
If there are multiple arguments, the result is
ordered by the first argument first, followed
by the second. 
For example, \code{UNION}
simply concatenates the two input dataframes in order, 
while \code{CROSS-PRODUCT} preserves
a nested order, where each tuple on the left 
is associated, in order, 
with each tuple on the right, 
with the order preserved.

\papertext{
We succinctly describe the new operators \rebuttal{as well as highlight
any deviating semantics of \code{GROUPBY} and \code{WINDOW}} and leave
detailed semantics to our technical report~\cite{petersohn2020towards}.
The output schema
for most other relational operators
can be carried over
from the inputs 
(indicated as {\em \rebuttal{static}} in Table~\ref{tab:df-algebra}).
}

\techreport{

Note that languages choose different approaches to inferring the schema 
after a \code{TRANSPOSE} with important implications for usability. 
For example, in R, a \code{TRANSPOSE} with heterogeneous $D_n$
ends up coercing everything to \textbf{string}, which may make it impossible to
apply another \code{TRANSPOSE} and yield a dataframe equivalent 
to the original $D_n$. In Python, everything is coerced to \code{Object}, which has typing
information embedded at runtime, so the schema induction function can
always recover the original $D_n$ after two transposes.}

% As a sidenote, this makes
% dataframes as defined here somewhat less attractive for operational settings like
% scheduled ETL jobs or streaming systems. \devin{I'm not sure I agree with this last sentence}
% \jmh{Somewhere we should point out that the presence of 
% schema induction functions means that the output 
% schema of certain operators like \code{TRANSPOSE} can be defined 
% to be null, meaning 
% they are unknown in the absence of a specific array of data. This
% can lead to type errors in algebraic expressions based on the values 
% of data in the inputs, which is impossible in either
% relational or linear algebra.
% It is no surprise that dataframes are most natural in languages
% like Python and R that support dynamic typing and runtime type errors.
% As a sidenote, this makes dataframes as defined here somewhat less
% attractive for operational settings like scheduled ETL jobs or streaming systems.}

\stitle{Transpose.} 
\code{TRANSPOSE} interchanges rows and columns, 
so that the columns of the
dataframe become the rows, and vice-versa.
Formally, given a dataframe $DF = (A_{mn}, R_m, C_n, D_n)$,
we define $\mbox{\code{TRANSPOSE}}(DF)$ to be a dataframe 
$(A_{nm}^T, C_n, R_m,$ $null)$,
where $A_{nm}^T$ is the array
transpose of $A_{mn}$. Note that
the schema of the result may be
induced by $S$, and may not be
similar to the schema of the input.
\code{TRANSPOSE} is useful both for 
matrix operations on homogenous dataframes,
and for data cleaning or for presentation 
of ``crosstabs'' data.
In step C2 in our example in Figure~\ref{fig:example}, 
the table was not oriented
properly from ingest, 
and a transpose was required to give us the desired
table orientation.

\techreport{
In \pandas and other dataframe implementations, 
it is possible to
perform many operations 
along either the rows or 
columns via the \code{axis}
argument. 
Instead, to minimize redundancy,
we define operators on collections
of rows, as in relational algebra,
and enable operations across 
columns by first performing a 
\code{TRANSPOSE}, applying
the operation, and then a \code{TRANSPOSE}
again to return to the original orientation.
That said, performing
\code{TRANSPOSE} can be expensive
(as we will see in Section~\ref{sec:modin}),
so one of our goals will be to postpone
performing it or avoid it entirely.
Moreover, given the presence of
\code{TRANSPOSE} in the algebra, we need to be prepared to handle
dataframes that are not only extremely high in cardinality (``tall'')
but also extremely high in arity (``wide'').

In the algebra defined above, we define
operators only on collections of rows, as in relational algebra, allowing \code{TRANSPOSE} to toggle
the axis of application of the operators.
Operations along the columns require a \code{TRANSPOSE}, application of 
the desired operator,
and a \code{TRANSPOSE} again to return to the
original orientation.
With this flexibility, operators
on the dataframe can be performed along either the columns or
the rows.

The asymmetry of 
row and column types in the relational model makes 
\code{TRANSPOSE} impossible to define for relations with non-homogeneous
column domains (for which the sets in $D_n$ differ):
there is no data-independent way to derive a relational 
output schema for \code{TRANSPOSE} from the input schema. In the 
dataframe data model, the data-dependent schema induction function
provides an output schema.

\code{TRANSPOSE} can also be extremely computationally expensive
depending on the system architecture and partitioning.
In its implementation, it will often
be important to postpone the calculation of \code{TRANSPOSE}
until the last possible moment because of the associated computation
costs. 
Moreover, given the presence of
\code{TRANSPOSE} in the algebra, we need to be prepared to handle
dataframes that are not only extremely high in cardinality (``tall'')
but also extremely high in arity (``wide'').
}

\stitle{Map.} 
% \jmh{I'm surprised that you defined Map to preserve the arity. I would have thought
% it could produce an arbitrary output row type! Also somewhat
% restrictive that Map is 1 row in and 1 row out. All of this will
% be a surprise to people familiar with Google MapReduce. You may want
% to rename or to comment.}
% \devin{The reason I defined it this way was to avoid everything becoming
% a map. It is easy to imagine a map operator that can produce an
% arbitrary output row type becoming a catch-all and replacing
% AUGMENT and FLATTEN as well. I tried to keep the definition as
% strict as possible to keep a clear distinction between operators
% that modify the R\_m and C\_n and those that do not. With that said,
% it could work if we redefine MAP to allow the user to specify an output R\_m and C\_n,
% but this complicates things and the current iteration of MAP is simple.}
The map operator takes some function {\em f}
and applies it to each row individually,
returning a single output row of fixed arity.
The purpose
of the map operator is to alter each dataframe row
uniformly. 
\code{MAP} is useful for data cleaning
and feature engineering
(e.g., step C3 in Figure~\ref{fig:example}).
Given a dataframe $DF = (A_{mn}, R_m, C_n, D_n)$, the result of
\code{MAP(DF, {\em f})} is a dataframe $(A_{mn'}', R_m, C_{n'}', D_{n'}')$
with $f:D_n \rightarrow D_{n'}'$, where
$A_{mn'}'$ is the result of the function {\em f} as applied to
each row, $C_{n'}'$ is the resulting column labels, and $D_{n'}'$ is the resulting
vector of domains. Notice that in this definition, 
the number of columns ($n'$) {\em and} the column
labels ($C_{n'}'$) can change based on this definition, but they must be changed
uniformly for every row.
The vector of domains $D_{n'}'$
may, in many cases, be inferred from the type
of the function $f$. 

% \jmh{Should we allow the optional declaration of an output
% domain $D'_n$ as a convenience?} \devin{Sure, that would be fine with me,
% but I am not sure how to present optional arguments in the notation we are using}

\techreport{
Extended relational algebra supports map via the use of functions
in the subscript of projection operators (i.e., in the \code{SELECT}
clause of SQL). However, this
% For example, in a relational system,
% the equivalent of a \code{fillna} operation would
% likely require code refactoring after a schema change introduces an
% additional column (wherein the new column is explicitly named in the
% \code{fillna} equivalent), but no such refactoring is necessary if
% an additional column is added to a dataframe.
projection
syntax is linear in the arity of the relation, which is cumbersome 
for very wide schemas (e.g., after a \code{TRANSPOSE}).
In this definition, \code{MAP} is passed an entire
row as an argument so it can reason across columns
in a generic fashion without enumerating them, whereas 
SQL expressions (including UDFs) typically require specific fields
from the row as scalar arguments. For example, 
consider a transformation that needs to ensure the values in 
all \textbf{float}-domain columns 
in a given row sum to 1.0; a generic, reusable \code{MAP} function can 
normalize the value in each \textbf{float} 
field by the sum of the \textbf{float} 
fields in that row;
instead, a SQL expression
would have to be crafted specially for each schema. 
\jmh{I think there's maybe more to say
here but we need a type theory person to help!}
\jmh{On a less technical note, can we assume the reader knows
\code{fillna} and that they realize it's a macro over \code{MAP}?}
}

\stitle{ToLabels.} 
The \code{TOLABELS} operator projects one column out
of the matrix of data, $A_{mn}$, to be set as new row labels for the resulting
dataframe, replacing the old labels.
% \jmh{If it's multiple columns, we either need vector-typed labels
% in our definition of a dataframe,
% or need to say that the labels are concatenated in some way.
% The previous sentence only makes sense if we allow vector-typed
% labels. I think we need that to support RESET LABELS. Actually,
% you're going further and saying that labels have data types too, which a notational mess we have to account for in our definitions. Bummer.}
Given $DF = (A_{mn}, R_m, C_n, D_n)$ 
and some column label $L$, 
\code{TOLABELS($DF$, $L$)} returns 
a dataframe $(A'_{m(n-1)}, L, C_n', D_n')$,
where $C_n'$ (respectively $D_n'$) is the result of removing the label $L$ from $C_n$ (respectively $D_n$).
With this capability,
data from $A_{mn}$ can be promoted
into the metadata of the dataframe and referenced
by name during future interactions.

\techreport{
From a relational perspective, this operator is rather unusual
in that it converts data into metadata.
Dataframe users are interested in wrangling and cleaning data,
so operations that let them move entries 
between metadata
and data are popular and convenient to use.
In fact, \code{TOLABELS}
followed by \code{TRANSPOSE} 
is, in effect, promoting data values into column
labels, which is impossible using 
relational operators.
}

% \smacke{I'm lost after reading this entire paragraph.
% Skipping and moving to the next operator.}
% \jmh{It's interesting that you chose to define AS LABELS on rows.
% In some sense it's just fine in the relational model, inasmuch
% as relations don't have a notion of row labels, you're just
% choosing to annotate some columns as special. If you had done this wrt
% to columns (e.g. promote a row of values to be column names) then 
% you've clearly stepped outside the first-order logic basis of 
% relational languages by allowing the data to affect the metadata.
% I guess in both cases you're affecting some metadata, but it's a bit
% mysterious here because there's no relational notion of ``special
% columns'' other than keys. Maybe you should point out that AS LABELS 
% followed by TRANSPOSE is in effect promoting data values into column
% labels, which is definitely not possible via any combination of relational operators.}

\stitle{FromLabels.} \code{FROMLABELS} creates a new dataframe 
with the row labels inserted
into the array $A_{mn}$ as a new column of data at position 0 with a provided column label.
The data type of the new column starts as $null$ until it can be induced by
the schema induction function $S$.
The row labels of the resulting dataframe are set to the default label:
the order rank of each row (positional notation). Formally,
given a dataframe $DF = (A_{mn}, R_m, C_n, D_n)$ and
a new column label $L$ we define
\code{FROMLABELS($DF$, $L$)} to be a dataframe
$(R_m + A_{mn}, P_m, [L] + C_n, [null] + D_n)$,
where $R_m + A_{mn}$ is the concatenation of the row labels
$R_m$ with the array of data $A_{mn}$, 
$P_m$ is the positional notation values for all of the rows: $P_m = (0, ..., m - 1)$,
and $[L] + C_n$ is the
result of prepending the new column label $L$ to the column labels $C_n$.

%\input{groupby-and-window.tex}
%!TEX root = modin-vision.tex

\stitle{\rebuttal{GroupBy.}}
\rebuttal{As in relational algebra, 
our \code{GROUPBY} operator
groups by one or more columns, 
and aggregates 
one or more columns
together or separately.
Unlike relational algebra,
where aggregation must result
in atomic values, 
dataframes can support composite
values within a cell,
allowing a broader class 
of aggregation functions
to be applied.
One special function, \code{collect},
groups rows with the 
same grouping attribute values 
into separate dataframes 
and returns these
as the (composite) aggregate values.
\Pandas's \code{groupby}
function has similar behavior and applies
\code{collect} to the non-grouped attributes\techreport{,
coupled with an implicit \code{TOLABELS} call
that elevates the grouping attribute values to the row labels}.
%  as an argument,
% and is always followed by an aggregate. The key difference with relational algebra
% is {\em first-class support for multivalued aggregates}, going so far as to admit
% a particular aggregate, \code{collect}, that groups rows
% with the same grouping attribute value into separate dataframes and returns these
% as the aggregate values. 
We will use 
\code{collect} in our examples subsequently. 
% An optimized query
% planner will, of course, need to avoid 
% materializing the sub-dataframes when possible.
}

% \stitle{\rebuttal{GroupBy.}}
% \rebuttal{
% The \code{GROUPBY} operator in our proposed algebra is analogous to that
% of traditional relational database systems: it accepts a column name as an argument,
% and is always followed by an aggregate. The key difference with relational algebra
% is {\em first-class support for multivalued aggregates}, going so far as to admit
% a particular aggregate, \code{collect}, that groups rows
% with the same grouping attribute value into separate dataframes and returns these
% as the aggregate values. \code{collect} appears, for example, in the pivot query
% plans of Figures~\ref{fig:pivotplan} and~\ref{fig:pivotplans}. An optimized query
% planner will, of course, need to avoid materializing the sub-dataframes when possible.
% }

\stitle{\rebuttal{Window.}}
\rebuttal{
\code{WINDOW}-type operations are largely analogous to those used
in recent SQL extensions to RDBMSs like PostgreSQL and SQL Server.
The key difference is that, in SQL, many windowing functions such
as \code{LAG} and \code{LEAD} require an additional \code{ORDER BY}
to be well-defined; in dataframe algebra, the inherent ordering
already present in dataframes makes such a clause purely optional.}

\techreport{
\code{FROMLABELS} is the opposite of 
the \code{TOLABELS} operator, and the two of these give the
user complete control over moving data to and from the
dataframe's labels. This allows users to apply operators
on the dataframe's metadata (specifically the row labels), 
which is particularly useful for
operators like \code{JOIN} and \code{GROUPBY}. 
Conceptually,
this operator also allows the positional 
notation of the dataframe to be treated as data if
multiple \code{FROMLABELS} are chained together. 
\techreport{However, because the order is immutable,
it is impossible to update the order of the dataframe
directly in this way. 
Despite providing the ability to promote data to row
labels (named notation), 
it is impossible in this algebra to
promote data to positional notation.
If the users wished to reorder the
data, they may \code{JOIN} with another dataset
with a specific order 
or \code{SORT} based on some column(s).}

From a relational point of view, \code{FROMLABELS} enables
the capability to push metadata into the data to be queried
and operated on. 
Thanks to this operator and \code{TOLABELS}
specifically, column and row labels must be of type
$\Sigma^*$ so that these operators make sense.
\code{FROMLABELS} also has some interesting interaction
with the schema induction function $S$, where labels
can be interpreted as any type in $Dom$ when they
are added to the data via \code{FROMLABELS} and then operated on. It
is important to point that out here in the definition, but we leave
the enumeration of the nuances of this interaction to future work.
}
\techreport{
\subsection{Algebra Examples}\label{sec:algebra-example}
}
\papertext{
\subsection{Pivot Case Study}\label{sec:algebra-example}
}
To demonstrate the expressiveness of the algebra above, 
we show how it can
\rebuttal{be used to elegantly and succinctly
express \code{pivot}, which is particularly
challenging in relational databases due to the need
for relations to be declared schema-first~\cite{wyss2005formal,cunningham2004pivot}. 
\techreport{The flexible schemata inherent in the dataframe data model
enables a succinct description of pivot.}
% We will demonstrate
% how the lazy schema described in Section~\ref{sec:dataframedef}
% enables us to perform the 
% data and metadata transformations and determine the schema
% {\em post hoc}, removing a significant challenge
% that is common in relational implementations.
}
% be used to capture a range of \pandas functions. 
%

\techreport{
To start off, many \pandas functions
provide essentially identical functionality
to dataframe operators, e.g., \code{sort\_values} for \code{SORT},
\code{merge} for \code{JOIN}, 
\code{groupby} for \code{GROUPBY}, 
\code{append} for \code{UNION},
\code{reset\_index} for \code{FROMLABELS},
and \code{set\_index} for \code{TOLABELS}.
The function
\code{transform} is a special case of \code{MAP} 
that applies a fixed function to each value 
within a row,
thereby preserving the input arity,
while \code{apply} is another special case
where a fixed function is applied on a per-row-basis
to combine values across multiple columns 
to generate a new column. 
}

\techreport{
A number of \pandas functions correspond to 
dataframe operators, with specific UDFs.
As examples for \code{WINDOW},
\code{cummax} computes the cumulative max of values
for one or more columns, \code{diff} takes the difference between
elements in a column and preceding values,
and \code{shift} shifts rows down to align with a new row label,
maintaining the order of the data. 
Likewise, for \code{MAP}, 
\code{fill\_na} converts 
all null values to another value,
\code{isna} replaces 
each value with a boolean based on whether or not
they are null, and 
\code{str.upper} converts all the string values to upper case.
In fact, \pandas has many functions that 
implement string and date-time transformations.
}

\techreport{
Finally, there are several \pandas
functions that are compositions of 
dataframe operators.
We list a few examples below,
with informal descriptions on 
how they may be rewritten using
the algebra.
}%

\techreport{
The \code{agg[`f1',`f2', ...]} function in \pandas
computes aggregate functions \code{f1}, \code{f2}, ..., 
for each of the columns individually, with the resulting dataframe
containing one row per aggregate, 
i.e., the first row corresponds
to the \code{f1} aggregates, 
the second to the \code{f2} aggregates, and so on. 
This function can be rewritten using one \code{GROUPBY} operator 
per aggregate function to produce a single row
corresponding to the aggregates, followed
by a \code{UNION} to append these rows to each other in
the order the aggregates are listed.
Another approach is to perform a \code{TRANSPOSE},
then a \code{MAP} to compute all the necessary aggregates,
one per column, followed by another \code{TRANSPOSE}
to bring the result to the right orientation.
}

\techreport{
% \removed{
The \pandas function \code{target.reindex\_like(reference)} 
supports changing a given
dataframe (the \code{target}) by reordering its rows and columns to match those of another dataframe (the \code{reference}). 
This operator is useful for aligning two dataframes
for comparison purposes.
One way to express this function using dataframe operators
would be to first \code{FROMLABELS} on both dataframes
to allow the row labels to become part of the data,
followed by a \code{INNER JOIN} between the two dataframes
on the row labels, 
with the reference as the left operand;
followed by a \code{MAP} to project out the reference 
dataframe attributes (leaving behind \code{reference}'s ordering).
Finally, \code{TOLABELS} can be used to move
the row labels back from the data. 
% }
}

%
%\begin{figure}
%    \centering
%    \includegraphics[width=0.45\textwidth]{pivot_figure.png}
%    \caption{%
%        Pivot Table. This figure is the same example presented in
%        \cite{cunningham2004pivot}, and demonstrates pivoting over two
%        separate columns, ``Month'' and ``Year''.
%    }
%\label{fig:pivot}
%\end{figure}
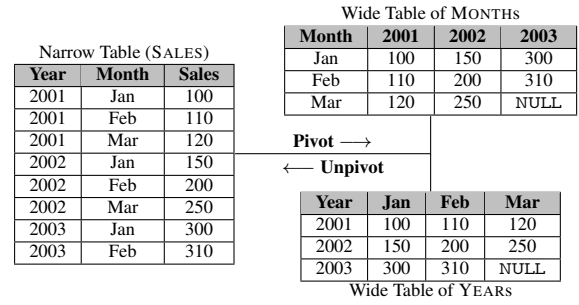
\begin{figure}[t]
\scriptsize
\centering
\begin{tikzpicture}
  \path
    (2.6,0.5) coordinate (A) node[above, inner sep=0] {
    \begin{tabular}{| c | c | c | c | }
	    \multicolumn{4}{c}{Wide Table of \textsc{Month}s} \\ \hline
        \rowcolor[HTML]{C0C0C0} 
    	\textbf{Month} & \textbf{2001} & \textbf{2002} & \textbf{2003} \\ \hline
    	Jan & 100 & 150 & 300 \\ \hline
    	Feb & 110 & 200 & 310 \\ \hline
    	Mar & 120 & 250 & \code{NULL} \\ \hline
    \end{tabular}
    }
    (2.6,-0.5) coordinate (B) node[below, inner sep=0] {
    \begin{tabular}{| c | c | c | c | }
        \rowcolor[HTML]{C0C0C0} \hline
    	\textbf{Year} & \textbf{Jan} & \textbf{Feb} & \textbf{Mar} \\ \hline
    	2001 & 100 & 110 & 120 \\ \hline
    	2002 & 150 & 200 & 250 \\ \hline
    	2003 & 300 & 310 & \code{NULL} \\ \hline
	    \multicolumn{4}{c}{Wide Table of \textsc{Year}s}
    \end{tabular}
    }
    (0,0) coordinate (C) node[left, inner sep=0] {
	    \begin{tabular}{| c | c | c | }
	        \multicolumn{3}{c}{Narrow Table (\textsc{Sales})} \\ \hline
        	\rowcolor[HTML]{C0C0C0} 
	    	\textbf{Year} & \textbf{Month} & \textbf{Sales} \\ \hline
	    	2001 & Jan & 100 \\ \hline
	    	2001 & Feb & 110 \\ \hline
	    	2001 & Mar & 120 \\ \hline
	    	2002 & Jan & 150 \\ \hline
	    	2002 & Feb & 200 \\ \hline
	    	2002 & Mar & 250 \\ \hline
	    	2003 & Jan & 300 \\ \hline
	    	2003 & Feb & 310 \\ \hline
	    \end{tabular}
    };
  \draw[-] (A)--(B) (2.6,0) -- 
	  node[above]{\textbf{Pivot} $\longrightarrow$}
	  node[below]{$\longleftarrow$ \textbf{Unpivot}}(C);
\end{tikzpicture}
\vspace{-5pt}
    \caption{%
        Pivot table example, reproduced
        from~\cite{cunningham2004pivot}, demonstrating
        pivoting over two separate columns, ``Month'' and ``Year''.
    }
\label{fig:pivot}
\end{figure}
The \code{pivot} operator elevates \rebuttal{a column of} data
into the column labels and creates a new dataframe reshaped
around these new labels (see Figure~\ref{fig:pivot}). 
\techreport{\rebuttal{The} \code{pivot} operator has been
described and implemented in relational systems~\cite{wyss2005formal, cunningham2004pivot}
but it is simpler to express in the algebra from Section~\ref{sec:df-algebra}.}

\begin{figure}[t]
{\vspace{-5pt}
\begin{tikzpicture}
\tikzset{
    mynode/.style={rectangle,rounded corners,draw=black},
    myarrow/.style={->, >=latex', shorten >=1pt, thick},
}  
   \node[mynode] (df) at (0,-0.75) {\code{DF}};
   \node[mynode] (year1) at (0,0) {\code{"Year"}};
   \node[mynode] (collect) at (1.3,0) {\code{collect}};
   \node[mynode] (groupby) at (1.3,-0.75) {\code{GROUPBY}};
   \node[mynode] (map) at (2.7,-0.75) {\code{MAP}};
   \node[mynode] (flatten) at (2.7,0) {\code{flatten}};
   \node[mynode] (tolabels) at (4.3,-0.75) {\code{TOLABELS}};
   \node[mynode] (year2) at (4.3,0) {\code{"Year"}};
   \node[mynode] (transpose) at (6.4,-0.75) {\code{TRANSPOSE}};
   \draw[myarrow] (df.south)-- ++(0,-.25) -- ++(.85,0) ->  (groupby.south);	
   \draw[myarrow] (groupby.south)-- ++(0,-.25) -- ++(1,0) ->  (map.south);	
   \draw[myarrow] (map.south)-- ++(0,-.25) -- ++(1.2,0) ->  (tolabels.south);	
   \draw[myarrow] (tolabels.south)-- ++(0,-.25) -- ++(1.7,0) ->  (transpose.south);
   \draw[myarrow] (year1.south) -> (groupby.north);	
   \draw[myarrow] (collect.south) -| (groupby.north);	
   \draw[myarrow] (year2.south) -| (tolabels.north);	
   \draw[myarrow] (flatten.south) -| (map.north);	
   \draw[myarrow] (transpose.east)-- ++(.5,0);	
\end{tikzpicture}
}
\vspace{-10pt}
\caption{Logical plan for pivoting a dataframe around the ``Year'' column
using the dataframe algebra from this section.}
\label{fig:pivotplan}
\vspace{-10pt}
\end{figure}
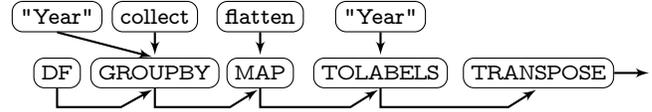

\rebuttal{Since} there is no need to know the names of the new columns 
\rebuttal{or the resulting schema} \emph{a priori}, 
\rebuttal{a pivot can be expressed concisely in dataframe algebra
as a combination of four operators in the plan shown
 in Figure~\ref{fig:pivotplan}.}
%\begin{lstlisting}[caption={Dataframe algebra expression for pivoting a dataframe around the ``Year'' column.},captionpos=b]
%MAP(
%    TOLABELS(
%        TRANSPOSE(
%            GROUPBY(DF, "Year")
%        ),
%    "Year"),
%flatten)
%\end{lstlisting}
Recall that it is possible to elevate data to the column labels by
using \code{TOLABELS} followed by \code{TRANSPOSE}. In this case, the \code{TOLABELS}
operator would be applied on the label of the column being pivoted over,
\rebuttal{\code{"Year"} in this example}. 
\rebuttal{After this step,
we perform a \code{GROUPBY} on the pivoted attribute, \code{"Year"}
with a \code{collect} aggregation applied to the remaining attributes to produce
a per-\code{Year} dataframe as a composite aggregated value.
This aggregated value is manipulated 
by a} \code{MAP} operator with
a function that flattens the grouped data into the correct orientation. 
This
results in a table pivoted around the attribute 
selected for the \code{TOLABELS}
operator. 
\rebuttal{
Notice in Figure~\ref{fig:pivot} that transposing the dataframe labeled
``Wide Table in Months'' results in the correct data layout for
the ``Wide Table in Years''. This is one example of how
\code{TRANSPOSE} can be exploited: cost models in dataframe
query optimizers can choose the more efficient pivot column
and \code{TRANSPOSE} at the end.}
% \agp{One more example?}
% \stitle{pivot.} \code{pivot} is an operator that modifies
% the shape and metadata of the dataframe to restructure how the
% data is consumed. Take the following example
% from prior relational algebra literature where the authors
% express pivot in a SQL system \cite{cunningham2004pivot}:

% \devin{Add table here}

% In the dataframe algebra defined in Section~\ref{sec:df-algebra}, the \code{TRANSPOSE}
% and \code{AS LABELS} operators removes the need to know the names of the
% columns a priori.
% The dataframe algebra composition of the \pandas
% \code{pivot\_table} operator on some dataframe $DF$ is as follows:

% \begin{lstlisting}
% # Promote Year column to labels
% DF_A = AS_LABELS(DF, "Year")
% # Year label becomes column names after TRANSPOSE
% DF_B = TRANSPOSE(DF_A)
% DF_C = GROUPBY(DF_B, "Month")
% # Transpose the rows in the context of the groups
% DF_D = MAP(DF_C, lambda row: TRANSPOSE(row)
% FINAL_DF = AS_LABELS(DF_D, "Month")
% \end{lstlisting}

\techreport{
To demonstrate the expressivity and power of
of this algebra, we demonstrate the real-world application of applying
it to \pandas.
In this section, we demonstrate how a few of the more
exotic \pandas operators can be written with
the algebra presented in Section \ref{sec:df-algebra}. We will illustrate that
there are operators which have a one-to-one mapping with the algebra
we described in the algebra, and that there are compositions of the
algebra operators. 

One-to-one mappings are shown in Table~\ref{tab:one_to_one}. \pandas has an
operator for each of the 

\begin{table}[]
\begin{tabular}{|p{0.125\textwidth}|p{0.09\textwidth}|p{0.2\textwidth}|}
\hline
\rowcolor[HTML]{C0C0C0} 
\textbf{Algebra Op}   & \textbf{Pandas Op} & \textbf{Pandas Op Description}                                             \\ \hline
                      & fillna             & Convert null values to another value                                       \\ \cline{2-3} 
\multirow{-2}{*}{MAP} & isnull             & Determine if elements are null                                             \\ \hline
TRANSPOSE             & transpose          & Exchange the columns and row                                               \\ \hline
AS\_LABELS            & set\_index         & Set the dataframe row labels using existing column(s)                      \\ \hline
RESET\_LABELS         & reset\_index       & Insert the row labels into the dataframe and set row labels to the default \\ \hline
\end{tabular}
\caption{
\pandas operators that directly map to algebra operators.
}
\label{tab:one_to_one}
\end{table}
}

\techreport{
Not all of the more than 200 \code{pandas.DataFrame} methods
map one-to-one to the algebra operators we have described. We now
describe some of the more interesting and complicated \pandas
operations and how they can be written in the algebra
we have defined.
}

\techreport{
\subsection{Extensions to the Formalism}\label{sec:dataframe-extensions}
\jmh{I feel like what's left here is too thin to bother with. Just say there's more in the tech report and drop the multiple labels (and hence this entire subsection).}

\todo{Make this a part of the original definition.}
Our data model so far is quite simple. 
We now describe a few additional extensions 
for our data model
that do not provide any additional expressive power,
but make certain operations more convenient.
\papertext{We describe one extension here,
and defer the rest to our technical report.
Our technical report
describes extensions to support
(i) label types, (ii) null values in labels,
as well as (iii) errors due to type mismatch detected
at runtime.}

\stitle{Multiple label columns.}
The data model can, optionally, have multiple row
label columns or multiple column label rows. 
Often, these are presented in a hierarchical
or nested manner in \pandas.
As an example, in a dataframe
tabulating sales, 
we could have two row label
rows that are nested, with the first (external) row label row
corresponding
to the years,
and the second (internal) row label row
corresponding to the quarters
within each year.
In our representation, we can simply capture
this by repeating the external row label values,
and combining the row label columns to give a single 
composite value, as shown below:
\begin{align*}
\small \left.
\begin{array}{lc||r}
2017 & Q1  &  \\
 	 & Q2  & \\
 	 & ... & A_{mn} \\
2018 & Q1  &   \\
	 & ... & 
\end{array}
\quad \implies \quad
\begin{array}{l||c}
	(2017, Q1) 	&  \\
	(2017, Q2)  & \\
 	... 		& A_{mn}\\
	(2018, Q1)  &  \\
 	... & 
\end{array}
  \right.
\end{align*}
}

\techreport{
\stitle{Label flexibility and types.}
Row labels can have a predefined type 
or domain from {\em Dom}---this type 
can be recorded separately and 
used to augment the schema $D_n$
when performing an \code{AS LABELS} operation,
thereby avoiding having to induce it using $S$.
Due to the symmetry between columns and rows,
column labels also have this constraint.
Additionally, labels can have duplicate values or be null;
so labels are not like primary keys. 

We finally define another notion
that will come in handy in future sections:
a {\em dataframe-like} system is one that supports
some, but not all dataframe properties as defined
in the data model and algebra above.
For example, a dataframe-like system might
support unordered weakly-typed relations,
with queries being composed incrementally
over the course of many statements. 
We return to this notion and provide some example
systems in Section~\ref{sec:ext}.

\subsubsection*{Workflow Definitions.}
We now briefly introduce 
some terms that will allow us
to describe how dataframes are manipulated 
during a data analysis workflow. 
% We now describe the definitions of different
% components of a given dataframe workflow. We will refer
% to these terms in our discussion of the algebra and future
% work sections.

\stitle{Operator.}
A {\em dataframe operator}, 
or simply an {\em operator},
is an atomic dataframe processing step
that
takes multiple
dataframe arguments and 
returns a dataframe as a result.
We will describe the operators in the context
of the dataframe algebra in Section~\ref{sec:df-algebra}.
% A dataframe operator is a single
% API call on or with a dataframe. This follows the SQL definition
% of an operator, see Section \ref{sec:df-algebra} for the operators
% exposed in dataframe APIs.

\stitle{Statement.} 
A {\em dataframe statement}
is an expression 
composed entirely of dataframe operators 
and is the unit of interaction between
the user and the system. 
In a notebook environment, a statement
corresponds to a single cell;
each of which is executed one at a time,
as we saw in Figure~\ref{fig:example}. 
In an interpreted environment
(e.g., iPython), a statement is a single block of code.

% A dataframe statement is the smallest unit of 
% submission a user can
% provide to the system and contains one or more operators.
% This presents itself in different ways depending on how
% the user is interacting with the dataframe system. In an interpreted
% environment (e.g. iPython), this will be a single block of code.
% In a notebook environment, a statement is a single cell. We define
% a statement in this way to draw parallels to a single SQL query statement
% in the way it is interacted with, however a statement is not a full
% query.

\stitle{Query.}
A sequence of statements 
chained together form a {\em dataframe query}.
Following variable references, a query
can be represented as a DAG of
operators and dataframes,
with the input dataframes 
at the leaves, and the queries as the root(s). 
A dataframe query is analogous to
a SQL query, but it is composed incrementally
across many statements.

% \stitle{Query.} A dataframe query is a series of statements that are
% composed to answer a larger overall question. If we consider the 
% data\-frame system as a DAG
% of operators and their dependencies, a query ends at each of the
% leaf nodes. A dataframe query is the equivalent of a SQL query.

\stitle{Session.} 
A {\em session} is a complete, end-to-end analysis workflow,
comprising one or more queries issued
across many statements.
A session begins when the user 
starts a notebook or interpreter environment
and ends when the user shuts down that environment.
% \jmh{Summing up what I think are the remaining concerns in the 
% formalism. What's down below is not intended as text for the paper,
% but I don't think you can put an itemized list into a comment. Meanwhile, we may still want this section in the end to keep things
% simpler up above and postpone boring complexity to here. We'll see.}
% \agp{We need to either expand on what's there below or skip. Ideally expand to cover at least what we promised under papertext above..  Right now it's just a set of notes.}
% \begin{itemize}
%     \item Row labels have types, like columns. This means that
%     column labels probably have to have types too for symmetry, or
%     transpose has to induce.
%     \item Labels are not like keys -- they may be missing, duplicated or NULL.
%     \item Dataframe expressions depend on data values for correctness
%     due to type induction; they can have ``runtime'' failures on
%     some data if type induction doesn't go as planned (e.g. you try
%     to do a SELECT "x < 10" on a categorical column.)
%     \item Need careful definitions of As Labels and Reset Labels once we fix the above.
    % \item Need a careful definition of AUGMENT, need to decide
    % if we want to define it as a complex macro or not. Maybe so--shows
    % the power of TRANSPOSE? \devin{At first it was convenience that led us to define the operator, but it does allow for some pretty nice display of the power of TRANSPOSE.}
% \end{itemize}

}

\techreport{
\begin{figure*}
    \centering
    \includegraphics[width=0.9\textwidth]{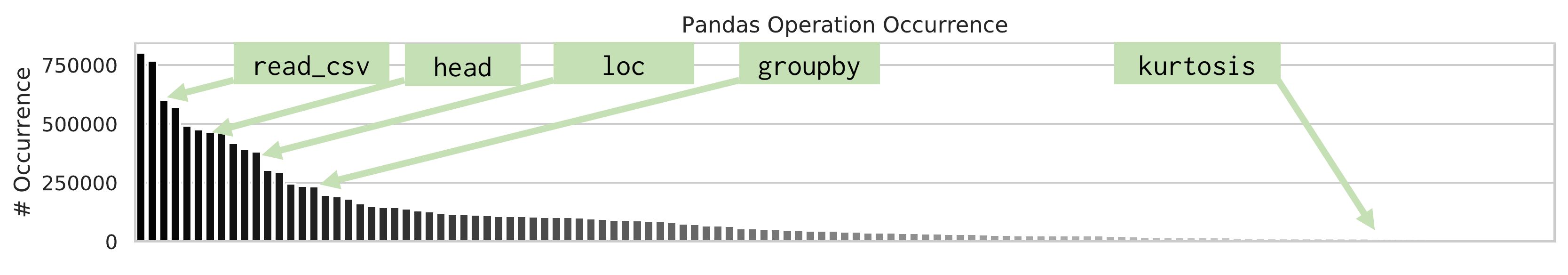}
    \caption{Pandas user statistics from GitHub dataset.}
    \label{fig:user_stat}
\end{figure*}

\subsection{Dataframe Usage Statistics}\label{sec:usage}

To study how dataframe users interact with the \pandas API, we analyzed a comprehensive dataset of 1 million Jupyter notebooks hosted on \hyperlink{github.com}{github.com} from Rule et al.~\cite{ucsd-dataframe-study}. 
Out of the 1 million Jupyter notebooks, about 40\% used \pandas.
We used the \code{jupyter nbconvert} module
to convert each 
notebook into to a 
python script, the \code{2to3} module 
to transform python 2 to python 3, 
and the python \code{ast}
module to parse and extract 
method invocation calls. 
We note that there may be some issues
in our extraction; for example, \code{.append} is both a python built in list method as well as a \pandas method. 
However, we expect our trends to largely
hold. 

We will focus on three questions to investigate how people work with \pandas. 

\stitle{What are some high-density functions used in interactive analysis?}
We studied the \textit{total} occurrence of each \pandas dataframe function in our data. 
The most notable ones are those
used 
to inspect the data (\code{plot, shape, head}), 
perform numeric aggregation (\code{mean, sum}),  
and
perform relational operations (\code{groupby, join}). 
It is worth noting that
the notebooks contained 
a lot of data modification operations
(both point queries as well as column and row queries) using \code{loc, iloc, drop, append}. 
Columns and index metadata inspection and manipulation are common as well with \code{index, columns}. 

\stitle{What kinds of functions are 
common in day-to-day usage?} 
We counted the number of files that each \pandas function
has occurred in. The occurrence 
measures the frequency of usage per analytic job.
The most commonly used functions
are those that create dataframes (\code{read\_csv, DataFrame}), inspect partial 
results (\code{head, shape}), visualizing result (\code{plot}), 
perform aggregation (\code{mean, sum, max}), 
perform point queries (\code{loc, ilo, ix}), add or remove data (\code{append, drop}), 
apply arbitrary user defined transformations (\code{apply}), and 
perform relational operations (\code{groupby, join}). 
It is worth noting that type-casting (\code{astype}) and direct access or manipulation of columns and index metadata, and underlying data storage (\code{columns, index, values}) are also high in the list. 

\stitle{Which functions are common used together?}
We also investigated the 
number of co-occurrence of functions 
in the same line of code. 
This typically involves
the user chaining \pandas functions 
together or calling them inside a single statement. 
For example, \code{df.dropna().describe()} is fairly common across our sample. 
It is also common for \pandas users 
to perform multiple operations in single execution cell. For example, \code{print(result = df["col1"].mean(), df["col1"].max())} prints a tuple of summary statistics. The popularity of chained or 
parallel invocation suggests 
opportunities for acceleration,
going beyond one operation at a
time to more complex queries. 
}
% Section 5
%!TEX root = modin-vision.tex

\section{data model challenges}\label{sec:datamodel}
% \dorx{New outline: \url{https://docs.google.com/document/d/13-pmh461S3Mj0WoVZ8l1zbN18nF7DI_luRzDRuUqe7U/edit?usp=sharing}}

\rebuttal{
Supporting 
the dataframe data model and algebra from
Section~\ref{sec:definitions} efficiently
motivates a new set of 
research challenges.
We organize
these challenges based on 
unique properties of dataframes,
and discuss their impact
on query optimization, data layout,
and metadata management. 
We first discuss the impact of flexible schemas.
%\devin{Expand for flow?}
% We draw on the distinctions between the
% dataframe and relational data models throughout.
% Here we outline the challenges in supporting the data model of the dataframe,
% as defined in Section~\ref{sec:dataframedef}. 
% These challenges
% are informed by our experience thus far
% with the implementation of a scalable dataframe system,
% \modin.
}

\subsection{Flexible Schemas, Dynamic Typing}
\label{sec:flexschemas}
%!TEX root = modin-vision.tex

\rebuttal{
Major challenges arise from the 
flexible nature of dataframe
schemas. 
Dataframes require more than data; 
as noted in Section~\ref{sec:dataframedef} 
they also require a schema to interpret the data.
In the absence of explicit types for certain columns, 
we must run the type induction function $S$, 
and the resulting parsing functions---both of which can be
expensive. 
Note that the type of a full column is required before 
we can parse the value of any cell in that column. 
Hence a major challenge for dataframes is 
to mitigate the costs inherent
in flexible schemas and dynamic types.

In database terms, dataframes are more like views than tables.
Programming languages like Python and R do not store data;
they access data from external storage like files or databases. Hence every time a 
program is executed, it constructs dataframe objects anew. 
Unfortunately, external storage in data science is often untyped.
Dataframe-friendly file formats like Apache Feather include explicit
schemas and pre-parsed data, but
most data files used in data science today (notably those in the 
ever-popular csv format) do not. 
% Fundamentally, a dataframe is by definition a programming
% language expression
% over the stored data.

Another source of dynamism arises from schema mutations,
e.g., adding or removing columns. These are first-class citizens of the
dataframe algebra, unlike in relational databases,
which relegate such operations to a separate DDL.
As such, they are
not only allowed, but are, in fact, frequent 
during data exploration
with dataframes, especially
during data preparation and feature engineering.
We consider
the challenge of efficient schema induction from three angles: 
rewriting, materialization, and 
query processing.
%\agp{revisit}
}
% When schema metadata such as the type 
% and number of columns
% are dynamic, 
% assumptions employed by traditional relational
% databases---for example, that queries can be statically
% checked for typing or semantic
% errors---no longer hold, 
% making it more challenging
% to perform query processing and optimization.
% Furthermore, by supporting \code{TRANSPOSE},
% dataframes promote rows to the same status as columns
% (and vice versa),
% giving rise to additional complications.
% These issues are
% % not special cases bolted
% % onto relational databases but are, in fact, 
% primary characteristics
% of dataframes. 
% It is therefore no accident that in our proposed dataframe algebra,
% the \code{TRANSPOSE} operation appears frequently as a component
% of longer, more complex chains of dataframe operations.
% Specifically, a major challenge for 
% supporting frequent schema mutation stems 
% from the necessity to work around the
% schema induction function $S$ if possible, 
% as it is an expensive operation.
% We explore this issue first.
% }

\rebuttal{
\subsubsection{Rewrite Rules for Schema Induction}
%Besides the difficulties introduced by the necessity of row metadata and the
%potential maintenance of such metadata, another challenge is deciding {\em when}
%to materialize such metadata.
%
%For example,
Due to their flexible schemas, dataframes 
support addition and removal
of columns as first-class operations, 
and at any point in time could
have several columns with unknown type. 
Certain dataframe operators need type
information, however---e.g. avoid attempting to 
\code{JOIN} two dataframes
on columns with mismatched types or using
a numeric predicate on a column with some strings.
The schema induction function, $S$, 
could be used to induce the requisite
typing information, but it is expensive, 
and must
be explicitly considered when
modeling cost for query plans.
Specifically, if certain columns are not operated
on, inferring their type via $S$
can be deferred to when they are 
first manipulated, and omitted entirely, if,
for example, they are dropped before ever
being accessed.

At least in some cases, 
schema inference rules might be able
to avoid the application of $S$ altogether. 
As one example, if ordered relational operations
are chained together, 
schema induction can be omitted between operations, 
suggesting the possibility
of employing rewrite rules to skip applying $S$. 
Another example involves
UDFs with known output types 
(e.g., a \code{MAP} with a 
UDF that always returns an integer).

In the case of operations which 
merely shuffle rows around (e.g. moving even-indexed
rows to the beginning of a dataframe, reordering), 
schema induction can be omitted entirely.
When filtering or taking a sample of a dataframe,
schema induction can be omitted if the type is already fairly
constrained and will not be additionally constrained
based on the sample. 
For example, if we
drop all rows with strings in a specific 
column, we may end up with that column having a restricted
type such as float or int, requiring special care.

While omitting or deferring schema inference is promising, 
additional complications arise 
from the fact that, in a dataframe system,
metadata {\em is} data (see also Section~\ref{sec:order_challenges}) 
that may itself be
queried by a user. 
In particular, it is common for users
to perform {\em runtime type inspections} 
as a sanity check.
As a result, the extra effort 
for eschewing or deferring schema induction may prove futile
if the user chooses to inspect types anyway. 
}

% Therefore, a truly intelligent
% optimizer could attempt to predict future user queries and 
% fuse schema induction together with other 
% operations that are type-agnostic (e.g., data movement
% or serialization/deserialization) while adding minimal overhead,
% the development of which we foresee to be a fruitful
% research direction.

%\subsubsection{Storage and Computation of ``Views''}
\rebuttal{\subsubsection{Reusing Type Information}}
\label{sec:matviews}

% Dataframes are flexible enough 
% to support both relational
% and linear algebra operations.
% Certain operations, such as the covariance operator, 
% places typing requirements
% on rows in addition to columns.
% This creates another dimension (literally) 
% over which to explore storage-compute tradeoffs.
% For example, a dataframe system could
% choose to materialize the row typing 
% information as part of a row schema 
% (as we briefly mentioned in Section~\ref{sec:dataframedef}),
% which could in turn be used 
% to avoid schema induction 
% following a \code{TRANSPOSE}.
% Indeed, without row-level schemas,
% performing two \code{TRANSPOSE}s in sequence
% results in the column schema $D_n$
% defaulting to an array of $\Sigma^*$,
% requiring explicit schema induction again.

\rebuttal{It is common to reuse a dataframe across 
multiple statements in a program. 
In cases where the dataframe lacks explicit types, 
it can be very helpful to materialize the results of 
both schema induction and parsing---both within the 
invocation of a program (internal state), and across invocations in storage.}

\rebuttal{Materialization of flexibly-typed schemas introduces
a new set of challenges. Both schema induction and parsing can be
a significant fraction of the cost of processing. 
This raises optimization choices for materialization: 
we can cache the results
of $S$ (for one or more columns), 
and additionally we can cache the results 
of parsing functions (in principle, at a granularity down to the cell level). 
For complex multistep dataframe expressions, 
we can choose to make these decisions at each operator
in the pipeline that introduces a dynamically-typed column. Hence the 
optimization search space is large.
Moreover, the workload of ``queries'' is different from 
traditional materialized view settings---languages
like Python are more difficult to analyze statically than SQL, 
and we can expect usage patterns to differ
from databases as well (Section~\ref{sec:interactivity}).}

\rebuttal{In some cases, it is reasonable to expect 
that a programmer will want to declare the 
types of the dataframe explicitly---e.g.,
an expression like \texttt{df\_t = \code{TRANSPOSE}(df, [myschema])} where \texttt{myschema}
is an array of type names for the columns.
In this case, there is no need to run schema induction. 
In a loosely-typed language like Python, 
\texttt{myschema} can be an arbitrary expression returning an array of strings.
For example, it might read a list of type names from a very large
file with the same number of rows as \texttt{\code{TRANSPOSE}(df)}. 
Alternatively, the dataframe \texttt{df} itself might have 
``row types'' stored as strings in the $i$'th column of the data, 
leading to an expression like
\texttt{df\_t = \code{TRANSPOSE}(df, df[i])}.} 
\agp{I think we need to add a bit here about why we may want to materialize
schemas prior to transposes.}

% \jmh{The next paragraph is channeling Stephen Macke's original discussion. I'm not sure I like this point though ...
% it's really not much like view maintenance, it's a more general return-type-induction problem. If we wanna talk about
% view maintenance for external files, we could, but that's also kinda yuck.}
\rebuttal{View maintenance has a role in the dataframe context, with new challenges for type induction.
The most direct use is in delta-computation of expressions that have the effect of ``adding'' rows to their inputs.
For example, consider a \code{MAP} operator with a data validation function: for each column it returns the input if it
passes a validation test, else it returns an error message in that column.
The new rows may all respect the constraints
of the types of the input dataframe, or some new rows could break those constraints---e.g. a string-typed error message
appearing in a column of numbers. In both cases, we'd like the type induction to take advantage of the work done to 
induce a schema for the input, and differentially decide on a schema for the output. Note that these issues get more subtle as 
the type system gets richer---e.g., consider an input with a column of type \emph{percent} 
that is passed into an arithmetic \code{MAP} function---the output may be statically 
guaranteed to be numeric, and for a given dataframe may or may not still be of type \emph{percent}.}

\rebuttal{Regardless of the source of the schema---whether it be induced, 
stored externally, or stored within the data itself---any 
implementation should assume that the metadata for a dataframe 
could be expensive to compute, and potentially very large. 
Storage and computation
of this metadata can have significant overheads, 
and methods for ameliorating those costs 
will be central to scalable dataframe
research.}

\rebuttal{\subsubsection{Pipelining Schema Induction in Query Plans}}
\rebuttal{When applying $S$ and the parsing function 
to columns is unavoidable,
we may be able to reduce its cost by trying to fuse
it with other 
operations that are type-agnostic and lightweight 
(e.g., data movement
or serialization/deserialization) while adding minimal overhead,
the development of which we foresee to be a fruitful
research direction.

% Furthermore, if a full application of the $S$ operator cannot be avoided,
% its position within the query plan can have major performance implications.
For other operations, the position of $S$ within
the query plan
can have major performance implications.
Consider a \code{MAP} operation that 
is being applied to a column of strings.
If the \code{MAP} operation is 
relatively inexpensive (e.g., if it is measuring the string length), 
it may make sense to to skip type checking via schema induction before the \code{MAP} operation. 
Although
a type error (due to, e.g., the presence of an unexpected integer value) 
%\jmh{how is an integer value like '12' not a string in our context?}
%\devin{if the user was expecting all strings but there is a column of integers,
% that will throw a type error if they assume it is string but it is not.
leads to wasted effort,
it may be acceptable, as the
overhead paid by actual application 
of the \code{MAP} is not too high. 
On the other hand, a \code{MAP}
which performs heavy-duty 
regular expression parsing 
over long strings may delay error detection
unacceptably if schema 
induction is fused with the \code{MAP} application.
%\jmh{I actually don't understand these examples.}

Overall, the positioning of the schema induction
operator within the query plan,
by possibly fusing it with existing operators,
combined with schema 
induction avoidance and reuse 
as previously discussed,
is crucial for the development of a
full-fledged dataframe query optimizer.}

\subsection{Order and Equivalence}\label{sec:order_challenges}
\rebuttal{
Unlike relations, dataframes are ordered
along both rows and columns---and users
rely on this ordering for debugging and validation
as they compose dataframe queries incrementally.
This order is maintained as rows are transformed
into columns and columns into rows via \code{TRANSPOSE},
ensuring near-equivalence of rows and columns.
Additionally, as we saw in Section~\ref{sec:algebra-example}, 
row and column label metadata is
tightly coupled with the dataframe content,
and inherits the order and typing
properties. 
% Unlike in
% relational databases, 
% where metadata is largely distinct from the data,
% data and
% metadata are interchangeable within dataframes. 
% Data can be promoted into metadata via \code{TOLABELS},
% likewise metadata can be demoted
% into the data via \code{FROMLABELS}. 
In this section, we discuss
the challenges imposed by enforcing
order and the frequently changing 
schema across row and column labels and row/column orientation.
}

% \todo{
% \squishlist
% \item All about layout
% \item Users expect to see data organized in specific way (hence order)
% \item Frequently change the layout (both by changing ordering, as well as transpose) and move data to metadata \& back
% \squishend
% }

\rebuttal{
\subsubsection{Order is Central}
The order of a dataframe is determined by the
order of ingested data. 
For example, a CSV file ingested as a 
dataframe would have the same 
row and column order as the file. 
This ordering is crucial for the trial-and-error-based
interaction between a user and a dataframe system.
Users expect to see the rows 
in their dataframe stay in the same order
as they process it---allowing
them to validate and debug each step
by comparing its result to the previous step.
For example, to ensure that
a CSV file is ingested and parsed correctly,
users will expect the first few rows
of the dataframe to be the same
as those they would see when examining the CSV
file. 
To examine a dataframe, users
will either use the operator
\code{head/tail} to see the prefix/suffix 
or simply type the name of the dataframe 
for both the prefix and suffix
in the expected order.
Additionally,
operators such as \code{WINDOW} and \code{MAP}  
from Section~\ref{sec:df-algebra}
expect a specific 
order for the rows (\code{WINDOW}) and
columns (\code{MAP}).
since the UDF 
argument to these operators may 
rely on that order. 
Dataframes also support \code{SELECTION} and \code{PROJECTION} 
based on the position of the rows and 
columns respectively. 
}

\rebuttal{
To support order, current dataframe systems
such as \pandas
physically store the dataframe
in the same conceptual order as defined by the user.
Said differently, they do not embrace physical data independence.
}\moved{%
% This constraint is not necessary,
% and removing 
Physical independence may open up new optimization
opportunities,
recognizing that as long as
the displayed results preserve 
the desired order semantics
to the users, it is not necessary 
that all intermediate products 
or artifacts (unobserved by user)
adhere to the order constraint.
For example,
a sort operation can be ``conceptual''
in that a new order can be defined without
actually performing the expensive
sorting operation.
Likewise, a transpose
doesn't require the data to be reoriented
in physical storage unless beneficial
for subsequent operations; the transpose can be
captured logically
to reflect the new
orientation of the dataframe.
}

\rebuttal{To ensure correct semantics
while respecting physical data independence,
we must devise means to capture ordering
information, either tracked 
as a separate ``order column'',
if it is not implied via existing columns,
or recording as metadata that the dataframe
must be ordered based on one or more 
of the preexisting
columns. 
Then, the \code{ORDER BY}
on this ``order column'' or one of the existing columns
will be treated as an
operator in the query plan, and will only need to be
done ``on-demand'' when the user requests to view a result.
% For example,
% users will not see the
% results from the intermediate artifacts of a query
% that are not explicitly displayed, so those artifacts
% do not need to be ordered.
Additionally, since users
are only ever looking at the first and/or last few lines
of the dataframe, those are the only lines that are required
to be ordered; we discuss this further
in Section~\ref{sec:immediate-feedback}.}

\rebuttal{%
% \stitle{Physical data independence with order.}
% \todo{Consider cutting this as related work}
Extending physical data independence even further,
we can adapt other data representation
techniques from the database community,
optimized for dataframes.
This includes columnar
or row-column hybrid storage~\cite{abadi2013design},
as well as those from 
scientific computing~\cite{choi1992scalapack},
array databases~\cite{rusu2013survey}
or spreadsheets~\cite{bendre2015dataspread}.
Since dataframes are neither relations,
matrices, arrays, or spreadsheets, none
of these representations are a perfect fit. 
% \jmh{I feel that the below is a narrow case,
% and would describe it as simply ``an additional 
% representation to consider'' rather than ``a promising 
% representation for dataframes'' which sounds like
% we propose it as the common case.}
% \devin{promising => candidate}
Given that rows and columns are equivalent,
one candidate for dataframe representation
is as a collection of key-value pairs,
where the key corresponds to the
(row number, column number)
pair.
This representation
is especially effective when
the dataframe is ``sparse'', allowing
us to omit pairs where the value is null.
Then, \code{TRANSPOSE} conceptually 
swaps the row and column for each value:
$(column, row, value)$,
and can be recorded in metadata.
However, some operations become more expensive,
e.g. reconstructing a row for a \code{MAP}
operation requires a join.
% Alternatively, a system could maintain
% multiple indexes for row and column
% combinations\agp{I don't follow this multiple indexes business}, or maintain multiple
% different physical layouts, however
% the memory cost here is likely to be prohibitive. \agp{this is a bit abrupt. you're talking about one representation but now you're jumping to a hybrid.}
% A promising research direction is to consider a 
% hybrid approach, where subsets of the data may
% be stored in different layouts, optimized
% for the common access patterns of that
% particular subset. 
% Overall,
Automatically detecting and updating to the right representation
% and updating it 
over the course of dataframe query
execution will be a substantial challenge.}

% \devin{Propose to cut entire next paragraph}
\techreport{
\moved{%
Given a certain physical representation,
operations on dataframes,
from a relational perspective, often make use
of ordered access, e.g.,
editing the $i^{\textrm{th}}$ row,
as well as access
based on the row labels, e.g.,
filtering based on row labels (named notation) 
or row position (positional notation).
Because
selecting the $i^{\textrm{th}}$ physical row or projecting the $j^{\textrm{th}}$ physical column
will not necessarily correspond to selecting (resp.\ projecting) the desired {\em logical}
row (resp.\ column), additional metadata
that serves as the "order column" or "order row" must be maintained to facilitate
order-independence of the physical data.
Automatically
maintaining
indexes for this purpose can be beneficial.
Recent work has developed positional
indexing~\cite{bendre2018towards}, allowing ordered
access to be supported in $O(\log n)$,
in the presence of edits (e.g., adding or removing
rows).
Column stores take a different approach
to avoid expensive edits across columns,
instead
recording edits separately as deltas,
and periodically merging them back in~\cite{abadi2013design};
it would be interesting to investigate which approach is more effective for a given set of dataframe operations.
Similarly, for matrices, accesses often
happen in a row-major or column-major order,
and identifying the right indexes to efficiently
support them in conjunction
with relational-style accesses, is an important
challenge.
In particular, when a dataframe
has many rows and many columns,
we may need both row-
and column-oriented indexing.
}
}

% \todo{
% \squishlist
% \item Preserving semantics: as long as we can reconstruct the ordering, we are OK.
% \item If not implied already from a column (e.g., via an order by) we can explicitly capture ordering via a separate column
% \squishend
% }
\begin{figure*}[t]
\centering
\begin{subfigure}{.45\linewidth}
\centering
\begin{tikzpicture}
\tikzset{
    mynode/.style={rectangle,rounded corners,draw=black},
    myslownode/.style={rectangle,rounded corners,draw=black,fill=pink},
    myarrow/.style={->, >=latex', shorten >=1pt, thick},
}  
   \node[mynode] (df) at (0,-0.75) {\code{DF}};
   \node[myslownode] (groupby) at (1.5,-0.75) {\code{GROUPBY}};
   \node[mynode] (collect) at (1.5,0) {\code{collect}};
   \node[mynode] (month1) at (0.125,0) {\code{"Month"}};
   \node[mynode] (map) at (3.2,-0.75) {\code{MAP}};
   \node[mynode] (flatten) at (3.2,0) {\code{flatten}};
   \node[mynode] (tolabels) at (5,-0.75) {\code{TOLABELS}};
   \node[mynode] (month2) at (5,0) {\code{"Month"}};
   \node[mynode] (transpose2) at (6.6,-0.75) {\code{T}};
   \draw[myarrow] (df.east) ->  (groupby.west);	
   \draw[myarrow] (groupby.east) -> (map.west);	
   \draw[myarrow] (map.east) -> (tolabels.west);	
   \draw[myarrow] (tolabels.east) ->  (transpose2.west);	
   \draw[myarrow] (month1.south) -> (groupby.north);	
   \draw[myarrow] (collect.south) -> (groupby.north);	
   \draw[myarrow] (month2.south) -| (tolabels.north);	
   \draw[myarrow] (flatten.south) -| (map.north);	
   \draw[myarrow] (transpose2.east)-- ++(.5,0);	
\end{tikzpicture}
\caption{Original plan}
\label{fig:pivotslow}
\end{subfigure}
\hfill
\begin{subfigure}{.45\linewidth}
\centering
\begin{tikzpicture}
\tikzset{
    mynode/.style={rectangle,rounded corners,draw=black},
    myfastnode/.style={rectangle,rounded corners,draw=black,fill=green},
    myarrow/.style={->, >=latex', shorten >=1pt, thick},
}  
   \node[mynode] (df) at (0,-0.75) {\code{DF}};
   \node[myfastnode] (groupby) at (1.5,-0.75) {\code{GROUPBY}};
   \node[mynode] (collect) at (1.5,0) {\code{collect}};
   \node[mynode] (year) at (0.25,0) {\code{"Year"}};
   \node[mynode] (map) at (3.2,-0.75) {\code{MAP}};
   \node[mynode] (flatten) at (3.2,0) {\code{flatten}};
   \node[myfastnode] (transpose1) at (4.2,-0.75) {\code{T}};
   \node[mynode] (tolabels) at (5.6,-0.75) {\code{TOLABELS}};
   \node[mynode] (month) at (5.6,0) {\code{"Month"}};
   \node[mynode] (transpose2) at (7,-0.75) {\code{T}};
   \draw[myarrow] (df.east) ->  (groupby.west);	
   \draw[myarrow] (groupby.east) -> (map.west);	
   \draw[myarrow] (map.east) -> (transpose1.west);	
   \draw[myarrow] (transpose1.east) ->  (tolabels.west);	
   \draw[myarrow] (tolabels.east) ->  (transpose2.west);	
   \draw[myarrow] (year.south) -> (groupby.north);	
   \draw[myarrow] (collect.south) -| (groupby.north);	
   \draw[myarrow] (month.south) -| (tolabels.north);	
   \draw[myarrow] (flatten.south) -| (map.north);	
   \draw[myarrow] (transpose2.east)-- ++(.5,0);	
\end{tikzpicture}
\caption{Optimized rewrite that leverages sorted \code{Year} column}
\label{fig:pivotoptimized}
\end{subfigure}
\vspace{-8pt}
\caption{Alternative query plans for pivoting
a dataframe around the ``Month'' column using the algebra from Section~\ref{sec:df-algebra}.
TRANSPOSE is abbreviated as T.}
\label{fig:pivotplans}
\vspace{-15pt}
\end{figure*}
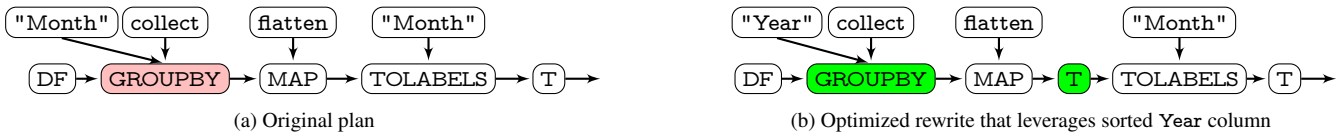

\rebuttal{
\subsubsection{Row/Column Equivalence}
\label{sec:rowcolumneq}
}
\rebuttal{
The presence of a \code{TRANSPOSE} operator in the
dataframe algebra
presents novel challenges in data layout and query optimization.
\code{TRANSPOSE} allows users to flexibly alter 
their data into a desired shape or schema that can be
parsed according to an appropriate schema, and 
queried using ordered relational operators.
}

\rebuttal{
To keep our data model and algebra compact, 
we have schemas only for columns, and
our operators are defined on ordered sets of rows.
By contrast, in \pandas and other dataframe implementations, 
it is possible to
perform many operations 
along either the rows or 
columns via the \code{axis}
argument.
Hence programs written in (or translated to) our algebra
are likely to have more uses of \code{TRANSPOSE} than dataframe
programs in the wild, to represent columnwise operations and/or
to reason about per-row schemas.}
% \rebuttal{
% As discussed in Section~\ref{sec:dataframedef}, our dataframe model assumes 
% a column-wise schema for a dataframe \texttt{df}. 
% If there is a desire to distinguish the types of the rows, then
% a new dataframe can be defined programmatically via \code{TRANSPOSE}, say \texttt{df\_t = df.\code{TRANSPOSE}()}. 
% By default, this requires running the type induction function $S$ on each of the columns of \texttt{df\_t}---i.e., the former rows 
% of \texttt{df}. 
% This is perhaps the most egregious example of the cost of metadata computation we discussed in Section~\ref{sec:matviews}.
% The issues raised there of materialized view caching and maintenance are critical to efficient computation of \code{TRANSPOSE}.
% }
% \moved{Instead, to minimize redundancy,
% we define operators on collections
% of rows, as in relational algebra,
% and}
\rebuttal{These operations are expressible logically in our simpler algebra
by first performing a 
\code{TRANSPOSE}, applying
the operation, and then a \code{TRANSPOSE}
again to return to the original orientation. Doing frequent physical reorganizations 
for these operations would be a mistake, however.
}
% \dlee{I don't think this paragraph is required, we should argue succinctly in the first paragraph that Transpose is important and leave it at that.
% - Transpose is frequent
% - Transpose is expensive}
% \rebuttal{
% \code{TRANSPOSE} allows users to conveniently leverage functions that are idiomatic
% for rows or columns, thus throwing off many data layout
% optimizations (e.g., columnar compression) specific to operators
% that are applied to rows/columns. 
% % \agp{abrupt -- how does this relate to the previous para?}
% % \agp{this is abrupt after the previous sentence.}
% For these reasons, \code{TRANSPOSE} is frequent in dataframe
% % Transpose operations that switch row and column axes are frequent in dataframe 
% workloads, since columnar operations are rewritten in terms of transposes in our 
% proposed dataframe algebra. 
% Changing the physical layout of the dataframe for 
% every transpose operation is expensive and unnecessary in most cases. One
% potential option is to perform transpose logically, by maintaining transpose axes
% information
% in the metadata and rewriting subsequent queries to leverage the metadata 
% information, instead of shuffling the layout upon every transpose operation.
% \agp{This part is ignoring what was discussed earlier wrt the key-value representation}

% \moved{
% Moreover, given the \rebuttal{frequent use of}
% \code{TRANSPOSE} in \rebuttal{typical dataframe workflows},
% we need to be prepared to handle
% dataframes that are not only extremely high in cardinality (``tall'')
% but also extremely high in dimensionality (``wide'').
% }
\rebuttal{The prevalence of \code{TRANSPOSE} in 
dataframe programs overturns many axis-specific assumptions made in traditional database storage.
Axis-specific data layouts like columnar compression are problematic in this context. 
Metadata management also requires rethinking, since dataframes are as likely to
be extremely
``wide'' (columnwise) as they are ``tall'' (rowwise).}
\rebuttal{
Both traditional and embedded RDBMSs typically limit the number of columns in a
relation (e.g., SQL Server has an upper limit of 1024
columns, or 30k columns using the wide-table feature)~\cite{raasveldt2019duckdb, sqlite2020hipp}. 
By applying \code{TRANSPOSE} on
a tall and narrow dataframe, 
the number of columns can easily exceed the millions 
in the resulting short and wide dataframe.
}

\rebuttal{
% \stitle{Query optimization.}
Dataframe systems
will need careful consideration to ensure that a \code{TRANSPOSE} call does not
break assumptions made by the data layout layer.
}
\rebuttal{
Given that the cost of performing a physical transpose will often
be high, 
one potential way to handle the data layout layer optimization
problem is to do a {\em logical} \code{TRANSPOSE} ``pull-up''. This would
delay or eliminate transpose in the physical plan as much as
possible since it will often destroy or render moot many
existing data layout optimizations. 
}

\rebuttal{
%As with most cost-based optimization, there are access patterns that admit
% exceptions to the rule. For instance, i
In certain cases, we may indeed want to consider optimizing
the \emph{physical} layout of the data given a \code{TRANSPOSE}
operator as a part of a query plan. 
This is in contrast with existing data
systems that create and optimize for a static data layout.
A physical transpose may help the optimizer match the layout to the
access pattern (e.g., matrix multiplication). A fixed data layout
is likely to have a significant performance penalty when
the access pattern changes.
Additionally, consider a case where \code{TRANSPOSE}
allows us more flexibility in query planning.
In the \code{pivot} case in Section~\ref{sec:algebra-example},
we observed that transposing the result of a pivot
is effectively a pivot across the other column. 
% This flexibility allows us to potentially do the \code{GROUP BY}
% on a sorted column, if it more cost effective
% to do the \code{TRANSPOSE} at the end.
Specifically, if we must pivot into the wide table with \code{Month}s as columns,
we can either use the original plan (Figure~\ref{fig:pivotplans}a) or  
one where we proceed as if the pivot is over \code{Year}, 
but then transpose the final result so that the \code{Month} attribute values
are used as column headers (Figure~\ref{fig:pivotplans}b). 
The latter plan will be faster if the optimizer leverages knowledge
about the sorted order of the \code{Year} column to avoid hashing the groups.
This is an interesting example of a
new class of potential optimizations within
dataframe query plans that exploit an efficient
\code{TRANSPOSE}. Because the axis transpositions are happening in query expressions,
the data layout becomes a physical plan property akin to ``interesting orders''~\cite{selinger1979access}
or ``hash teams''~\cite{graefe1998hash},  expanding the rules for query optimization.
}

\rebuttal{
\subsubsection{Metadata is Data (and Data is Metadata)}
A standard feature of dataframes is the ability to fluidly move values from
data to metadata and back. This is made explicit in the \code{TOLABELS} and 
\code{FROMLABELS} operators of our algebra, especially in combination with \code{TRANSPOSE}. 
These semantics cannot be 
represented in languages like SQL or relational algebra that are grounded in 
first-order logic; this is a signature of second-order logic, as explored in
languages like OQL~\cite{alashqur1989oql}, SchemaSQL~\cite{lakshmanan1996schemasql} 
and XQuery~\cite{chamberlin2001xquery}.
There is significant prior work on optimizing second-order operations like
the unnesting of nested data (e.g.~\cite{fegaras1998query, shanmugasundaram2001querying, wang2007optimization}).
A distinguishing aspect of our setting is that a dataframe operation like \code{TOLABELS} commonly generates a
volume of metadata that is dependent on the size of the data; this raises new challenges.
The closest prior work to our needs
is on implementing spreadsheet-style pivot/unpivot in databases (e.g., ~\cite{cunningham2004pivot,wyss2005formal});
this work needs to be generalized to the richer semantics of a dataframe algebra.
}
% \subsubsection{Metadata is Data (and Data is Metadata).}
% Dataframes have the unique capability to move data to metadata
% and back. This is available through the \code{TOLABELS}
% and \code{FROMLABELS} operators, where a column of
% data is elevated to the row labels in \code{TOLABELS},
% or the labels are inserted into the data with \code{FROMLABELS}.
% This capability provides extreme flexibility in manipulating
% and coercing data, but presents some new challenges in
% metadata management.
% }

\rebuttal{To address representational aspects, we could treat row labels
the way we treat primary keys in a relational database---by noting the sequence of label columns
in a metadata catalog. Some additional details arise in the support of positional notation:
invoking \code{TOLABELS($c_1$, ..., $c_n$)} removes the relevant columns from their positions, requiring
a recalculation of the positions of all labels to the right of $c_1$. This can be
handled by representing column order in dynamic ranked data structures like ranked B-trees~\cite{knuth1997art} or 
range min-max trees~\cite{navarro2014fully}. In terms of data access, we may want to efficiently process 
data columns without paying to access (dynamically reassigned) metadata columns, and vice versa. In this case, columnar 
layouts become attractive for projection. Alternatively,
labels can be moved into separate \emph{property tables}~\cite{cunningham2004pivot}, a form of ``vertical partitioning'' that does not rely on
columnar storage layouts.}

\rebuttal{Challenges arise in more complex expressions that include both \code{TOLABELS} and other operators--notably \code{MAP} 
and \code{TRANSPOSE}. In these cases, the number and types of \emph{columns} in the dataframe is data-dependent. This exacerbates the 
metadata storage issues discussed in the previous section, and brings up additional challenges.}

\rebuttal{In terms of query optimization, we now have a two-dimensional estimation problem: both cardinality estimation (\#rows) 
and \emph{arity estimation} (\#columns). For most operations in our algebra this would appear straightforward: even for \code{TRANSPOSE}, we
know the cardinality and arity of output based on input. The challenge that arises is easy to see in a standard data science ``macro'',
namely 1-hot encoding (\texttt{get\_dummies} in \pandas). This operation takes a single column as input, and produces a result table whose schema
concatenates the input schema with an (typically large) array of boolean-typed columns, one column per distinct data value of the input. Pivot presents
a similar challenge: the width of the output schema is based on the number of distinct data values in the input columns. In our algebra, these
macros can be implemented using \code{GROUPBY} followed by \code{MAP} and \code{TRANSPOSE}. The resulting arity estimation problem reduces to 
distinct value estimation for the input to \code{GROUPBY}.
Techniques like hyperloglog sketches~\cite{flajolet2007hyperloglog} come to mind to assist here. But
note that we need to compute these estimates not only on base tables that may be pre-sketched, but on intermediate results of expressions! In short, 
we need to do distinct value estimation for the \emph{outputs} of query operators---including arithmetic calculations (e.g. sums, products) and string manipulations (e.g. expanding a document into constituent words).}

\rebuttal{In some scenarios, arity \emph{estimation} is insufficient---we need exact numbers and labels of columns. Consider the example of performing a \code{UNION}
of feature vectors generated from two different text corpora, say wikipedia articles unioned with DBLP articles. 
Each text corpus begins as a dataframe with schema \texttt{(documentID, content)}. After a standard series of text featurization steps
(word extraction with stemming and stop-word filtering followed by 1-hot encoding), each corpus becomes a dataframe with a \texttt{documentID} column, and one boolean column for each word in the corpus. The problem is that the \code{UNION} needs to dynamically check for compatibility of the input schemas---it needs to first generate the full (large!) schema for each input, and compare the two. 
Even if we relax our semantics to an ``outer'' union, 
we want to identify and align the common words across the corpora. 
These metadata requirements seem to require two passes of the inner expression's data: one to compute and align metadata, 
and another to produce a result.
There are opportunities for optimization here to return to single-pass pipelining techniques, but they merit thoughtful investigation. This pipeline-breaking problem generalizes to any operator that reasons about its input schema(s), so it needs to be
handled comprehensively. 
\\}
\rebuttal{
In short, we expect that the fluid movement of large volumes of data into metadata and vice versa 
introduces new challenges for query processing and optimization in dataframes.}

\section{User Model Challenges}\label{sec:interactivity}
% \todo{Reorient sections}
% \todo{(1) incremental queries}
% \todo{(2) Debugging - cite something}

\rebuttal{
Unlike in SQL where queries are submitted {\em all-or-nothing}, 
dataframe users construct queries 
in an incremental, iterative, 
and interactive fashion.
Queries are submitted as a series of {\em \stmts} (as we saw in Figure~\ref{fig:example}), 
a few operators at a time, 
in trial-and-error-based \textit{sessions}. 
Users rely on immediate feedback 
to debug and rapidly iterate on these \stmts 
and frequently revisit results of intermediate \stmts 
for experimentation and composition 
during exploration. 
This interactive session-based 
programming model for dataframes 
creates novel challenges for overall system performance
and imposes additional constraints on query optimization. 
For example, operator reordering is often not beneficial
when the results are materialized for viewing after every \stmt.
At the same time, dataframe query 
development sessions are bursty, 
with ample think time between issuance of \stmts, 
and tolerant of incomplete results as feedback---as long
as the original goals of experimentation and debugging are met, 
offering new opportunities for query optimization.
In this section, we discuss new
challenges and opportunities in query optimization 
arising from the interactive and incremental
trial-and-error query construction of a typical
 user.
}

\subsection{Interactive Feedback and Control}\label{sec:immediate-feedback}
\rebuttal{
Dataframes are typically used in 
exploratory workloads, 
where interactive response times 
are crucial to providing a fluid user experience. 
Past studies have shown that latency 
in response times of greater than 500ms can lead
to fewer hypotheses explored and
insights generated during data exploration~\cite{liu2014effects}. 
As another example, 
for data preparation---often performed on dataframes---users 
often rely on system feedback 
to guide and decide what operations 
to determine their next steps~\cite{Heer2015}.
This feedback usually comes in the form
of a display output by the dataframe system
that contains a prefix or suffix of rows and columns,
as in Figure~\ref{fig:example}.
The need for frequent materialization 
of intermediate results of \stmts
to provide feedback to the user
% for feedback \agp{what is feedback here? is it for deciding what to do next? for debugging? we should clarify globally} 
makes it particularly 
difficult to satisfy the 500ms query 
latency requirement for interactivity.  
Fortunately, we can leverage two
user behavior characteristics to improve
interactivity:
that users spend time thinking between 
steps, and that the inspected intermediate 
results are typically restricted 
to a prefix/suffix of rows/columns is sufficient 
for debugging
and validation.}
\rebuttal{
\subsubsection{Intermediate Result Inspection \& Think Time}
\label{sec:opportunistic-evaluation}
Present-day dataframe systems such as \pandas
are targeted toward ensuring users
can inspect intermediate results for debugging and validation, 
so they operate
in an {\em eager} mode where every \stmt is evaluated as
soon as it is issued.
Program control is not returned to the user
until the \stmt has been completely evaluated,
forcing the user to be idle during that time.
However, there are many cases where users do not inspect the
intermediate results, or where results are discarded;
in such cases, the user is still forced to wait for each \stmt to
be evaluated. 
Moreover, users are either rewarded or punished
based on the efficiency of a query as it is written.

On the other hand, with the {\em lazy} mode of evaluation,
\rebuttal{which is adopted by some dataframe-like} 
systems~\cite{dask, Armbrust:2015:SSR:2723372.2742797} (See Section~\ref{sec:ext}),
control is returned to the user immediately,
and the system defers the computation until the user
requests the result.
By scheduling computation later,
the system can wait for larger query sub-expressions to be assembled,
leading to greater opportunities for optimization.
The downside of lazy evaluation is that
computation only begins when the user
requests the result of a query.
This introduces new burdens for users,
particularly for debugging, since bugs are not revealed
until computation is triggered.

For example,
consider two commutative operations \texttt{op1}, and \texttt{op2}. 
Say the user submits the \stmt \texttt{x = df.op1()}
followed by \texttt{y = x.op2()}. 
In eager evaluation, \code{x} will be fully materialized before
execution begins on \code{y}, even if \code{x} is never used again. 
Computing \code{y} could be done using 
\texttt{df.op2().op1()}, but it is often more beneficial
to use the materialized version of \code{x} instead.
In lazy evaluation, execution will be deferred until
explicitly requested, so the
expression that creates \code{y}
could be optimized to run \texttt{df.op2().op1()}.
The downside of this approach is that the user has to wait
until they explicitly request \code{y} before
they realize that there is a potential bug in \code{x}.
% For example, if the user composes a cross-product
% followed by a filter, this filter
% can be rewritten to be performed first.
% \jmh{be clear what you mean by a ``step'' - is it a part of an dotted chain expression, or a request for output to be computed (Shift-Return) or what...}

Furthermore, neither the lazy nor the eager mode
take advantage of the fact that the users
spend time thinking between steps:
the system is idle during think-time.
We can can leverage this time for computation,
allowing us to effectively achieve the benefits
of both paradigms.
% For smaller datasets or simple queries this can
% potentially lead to a longer overall runtime if there are
% periods of think time
% that exceed the runtime of the statements
% in the query.
% \techreport{In fact, most users of such systems
%  end up requesting the results after each operation,
%  essentially defaulting to the eager mode.}
While interactive latency is important 
to support immediate feedback, recent 
empirical studies have also shown 
that optimizations can be relaxed to account 
for user's long think time 
between operations in exploratory 
analysis~\cite{Battle2019}. 
We describe a novel \emph{opportunistic} query 
evaluation paradigm suitable for optimizing dataframes 
in an interactive setting.
}

\moved{
Like lazy evaluation, opportunistic evaluation
does not require the user to wait after each statement.
Instead, the system opportunistically
starts execution, while passing control back to users
with a pointer to the eventually computed dataframe (a ''future''),
which is asynchronously computed in the background.
We can then use system resources to compute
results in the background as users
are composing the next step.
Like eager evaluation, opportunistic evaluation
does not wait for users to complete the entire
query to begin evaluation.
However, when a user requests to view a certain
output, opportunistic evaluation can prioritize
producing that output over all else.
Opportunistic evaluation allows queries to be rewritten 
as new \stmts are submitted \rebuttal{(e.g., \texttt{df.op2().op1()})} 
to get to the requested
answer as fast as possible, taking into account
what is partially computed.
There are also new opportunities within opportunistic evaluation
to do speculation, 
where during 
idle time the system can start
executing \stmts that commonly follow previous ones.
Opportunistic evaluation also leads to new
challenges in sharing and reuse
across many query fragments whose computation
has been scheduled in the background (see also
 Section~\ref{sec:trialanderror}).
}

\rebuttal{
\subsubsection{Prefix and Suffix Inspection}\label{sec:prefix}
\rebuttal{
The most common form of feedback provided 
by dataframe systems is the tabular
view of the dataframe, as shown in Figure~\ref{fig:example}. 
The tabular view
serves as a form of visualization 
that not only allows users to inspect
individual data values, 
but also convey the structural information associated
with the dataframe. 
Structural information, especially as it relates to order,
is important for validating the 
results of queries that manipulate and
reshapes the dataframe. 
This tabular visualization typically 
contains a partial view of the dataframe,
displaying the first and last few rows of the dataframe,
accessed using \code{head}, \code{tail}, or other print commands.
% During dataframe
% query construction, table visualizations are frequently requested via the
% \code{head}, \code{tail}, or other print commands, as we saw in
% Section~\ref{sec:usecase}.
}
% \moved{\agp{There is a lot of repetition in the following lines 
% w.r.t the previous.
% The partial view stuff is repeated. The order is repeated. Once is plenty.}
% % As we saw in Section~\ref{sec:usecase}, 
% % building queries incrementally
% % through trial-and-error
% % typically involves printing
% % the first or last few rows of the dataframe,
% % via the \code{head}, \code{tail},
% % or other print commands.
% The \code{head} and \code{tail} commands,
% in \pandas, print the first
% or last $k$ rows (5 by default), respectively,
% while printing a dataframe displays
% both the first and last $k$ rows.
% When building queries through trial-and-error, 
% the user
% relies heavily on the order of the displayed data
% in judging the correctness
% of the intermediate
% query sub-expression,
% and deciding what to do next.
% When the user employs these printing commands,
% they expect the same $k$ rows to be displayed,
% in order.
% }

One way to give the
users immediate feedback
is to return
the output to the user as soon as these $k$
rows are assembled, computing the rest of the output
in the background using opportunistic evaluation.
This is reminiscent of techniques that
optimize for early
results~\cite{viglas2003maximizing,viglas2002rate}
for \code{LIMIT} queries~\cite{kim2018optimally},
or 
for representative tuple identification~\cite{singh2012skimmer},
but a key difference in dataframes is that order
must be preserved (so "any-k" result tuples will not suffice~\cite{kim2018optimally}), 
and there are many more blocking operators.
% \agp{I think we should "amp up" the differences.
% Omit special care etc. Say a KEY difference is that the order
% must be preserved and that there are many more blocking operators
% } 
% Therefore, techniques that try to find
% samples anywhere in the data that match query predicates
% as in Kim et al.~\cite{kim2018optimally},
% will not apply.
One starting point would be to design or select physical operator 
implementations that not just prioritize high output
rate~\cite{viglas2002rate},
but also preserve order, thereby ensuring
that the first $k$ rows will be produced as quickly as possible.
As an example, if only the first $k$ rows
of an ordered join were to be computed,
a nested loop join, with the result displayed
after $k$ rows are computed, might work well.
We can progressively process more portions of the
input dataframes until $k$ output rows are produced
in order: this may mean processing more than $k$ rows
of the inputs if there are very selective predicates.
Figuring out the right way to exploit parallelism
to prioritize processing the prefixes
of the ordered input dataframes 
to produce the ordered prefix of the output is likely
to be a substantial challenge.

Additionally, certain blocking operators will cause problems.
While returning the first $k$ rows following a
\code{TRANSPOSE}, especially when using columnar storage,
can be fairly efficient,
it may be hard to produce the first $k$ tuples
of a \code{GROUP BY} or \code{SORT}
without examining the entire data first. 
\techreport{(Indeed,
for small enough $k$, sorting may be faster
than $O(n \log{n})$, but still requires an $O(n)$ sequential scan.)}
% Optimizers in RDBMSs draw upon techniques such as predicate pushdown to reduce the amount of data processed for a query. 
% Unfortunately, due to the order requirement in dataframes, predicate pushdown is often not directly applicable to prefix/suffix computation. 
\code{SORT} is an obvious example where the top $k$ rows cannot be narrowed down a priori; 
a full scan is inevitable also for \code{GROUPBY} on non-clustered columns. 
That said, 
since the top and bottom $k$ rows are often the only results 
inspected for dataframe queries,
we may benefit from materializing additional intermediate results
or supporting indexes to retrieve them efficiently, without exorbitant 
storage overhead.
We could, for example, materialize the prefix and suffix
of a dataframe in original and transposed orientations,
or the prefix or suffix of the dataframe sorted by various
columns to allow for efficient processing subsequently.
These materializations could happen during think-time 
as discussed in Section~\ref{sec:opportunistic-evaluation}.
% , maintaining additional intermediate results and data structures for these regions \`{a} la incremental view maintenance can significantly speed up query processing without exorbitant storage overhead. What these data structures are and how to maintenance them in the ordered setting present interesting research opportunities.
We may also be able to exploit approximate query processing
to produce
the prefix/suffix early for blocking 
operators~\cite{ding2016sample+,agarwal2013blinkdb,park2018verdictdb,hellerstein1997online,zeng2014analytical}.
Since the tabular view is only a special form of visualization,
a rich body of related work from visualization on
how to allow users to quickly but approximately make decisions or perform debugging or validation, 
may be applicable~\cite{kim2015rapid,park2016visualization,macke2018adaptive,alabi2016pfunk};
however, the rich space of operators that goes beyond simple \code{GROUPBY} aggregation
will lead to new challenges.
Another interesting usability-oriented 
challenge is whether this tabular
view of prefixes or suffixes 
is indeed best for debugging---perhaps highlighting
possible erroneous values or outliers
in dataframe rows or columns that are not in the prefix
or suffix may also be valuable~\cite{Raman2001}.

% Recent work has also introduced
% the notion of guarantees
% across collections of aggregates,
% often optimizing for user tasks or
% perceptual
% guarantees \rebuttal{for visualizations}~\cite{kim2015rapid,park2016visualization,macke2018adaptive,alabi2016pfunk},
% e.g., ensuring that the
% ordering of the aggregates
% matches the correct ordering,
% even if the values of the aggregates are
% inexact,
% and these techniques
% may be a good fit to the dataframe setting.

% For example, a \code{GROUP BY} might correctly produce
% an approximate estimate
% of the first $k$ groups without
% filling in the aggregates, and a \code{SORT}
% might be able to produce an approximation
% of the top-$k$ after reading a large
% enough random sample\techreport{---however
% guarantees on correctness may be minimal}.
% \agp{See if we can amp up the novelty for group by and sort -- they
% are pretty traditional ops}
% \dorx{How's the new blurb?} \devin{it is great}
}

\techreport{    
\moved{
% \dorx{Move the rest of this section to TR?}
While it is well known that some aggregates like \code{MAX}
cannot be approximated~\cite{mozafari2015handbook},
even blocking operators such as \code{SORT} can return
early approximate results, using results
from the 1990s~\cite{raman1999online}.
There may be additional opportunities for approximation
if the user simply wants to inspect the approximate
structure of the result for debugging purposes,
especially in conjunction with prefix/suffix computation.
For example, we can provide the
overall structure of the output of
a pivot table computation
(displaying the row-wise groups and column headers),
without actually filling
in any of the aggregate values, and doing so
progressively.
Similar ideas of adding ``placeholder'' values
for in-progress tuples have been proposed in
streaming~\cite{raman2002partial}, web-database
hybrid~\cite{goldman2000wsq},
and crowdsourcing~\cite{parameswaran2012deco} contexts,
but not, as far as we can tell, for a group-by aggregation setting.
This idea could also be applied, 
for example, to \code{TRANSPOSE},
where the structure of the 
output dataframe is prepared first, with
the values filled in progressively.
}
% \devin{One paragraph tech-reported here}
\moved{
Other notions of approximation may also be valuable,
e.g., the incomplete/phantom notions in Lang et al.~\cite{lang2014partial},
wherein the result may contain additional rows
not present in the dataframe query result,
or rows that should be present, but are absent.
This could be valuable, for example, for expensive filters.

In fact, we could also exploit correlations~\cite{joglekar2015exploiting}
between the filtering attribute
and the other attributes in order to quickly approximate
the rows that might pass the filter and
quickly display them to the user, refining as
additional filter evaluations are performed.
}
}

%!TEX root = modin-vision.tex

\subsection{Incremental Query Construction}\label{sec:trialanderror}
% and incremental query construction
\rebuttal{
% \agp{I think this paragraph needs to be reorganized
% not around immediate feedback--which is the focus
% of the previous subsection---but around the ad-hoc incremental
% composition behavior. We should not mention sharing and reusing upfront---it is a technique that helps improve interactivity for this behavior. We can bring it up later.}
% Given the importance of providing immediate feedback to users when working with dataframs interactively, in this section, we further describe opportunities to speed up query execution by {\em sharing and reusing} across the session. 
% Users often build dataframe queries incrementally over the course of an interactive session through trial-and-error. They iterate on subqueries and assemble multiple query statements together to accomplish a larger task. As a result of this highly iterative process, query sub-expressions are often repeated multiple times. There are opportunities for shared computation via {\em multi-query optimization}, by computing many query sub-expressions at the same time, and for reuse via {\em materialization},
% by caching intermediate dataframe results and reusing to process subsequent query sub-expressions.
%
% challenge: 1) need frequent materialization 2) how to optimize across operators with frequent result mat required?
% opportunity: revisit and share subquery results; redundant computation
In addition to the challenges around 
satisfying the stringent latency requirement 
for immediate feedback discussed above, 
the need to frequently evaluate and display results for 
intermediate sub-expressions (i.e., the results of \stmts) 
over the course of a session
complicates query optimization,
as we saw in Section~\ref{sec:opportunistic-evaluation}.
While incrementally  
constructing dataframe queries 
over the course of an interactive session, 
users iterate on query sub-expressions through trial-and-error, 
frequently inspecting and revisiting intermediate results
to try alternate exploration paths. 
Such fragmented workloads limit the optimizations 
that can be applied to each sub-expression. 
However, since
user \stmts often build on others,
we can jointly optimize across these \stmts and resulting sub-expressions,
sharing work as much as is feasible.
Further, since users commonly return
to old \stmts to try out new exploration paths,
we can leverage materialization to avoid redundant 
reexecution.
We discuss these two ideas next.

% We can overcome this limitation by applying {\em multi-query optimization}. Sub-expressions in a session often operate on overlapping data or build upon a shared set of previous sub-expressions. This creates ample opportunities for sharing computation and intermediate results across queries.
% \dorx{another sentence on the specifics if we were so inclined. not necessary IMO.}

% In this section, we describe
% research challenges that leverage
% this session-based mode of operation to further
% optimize the user experience.
% We identify two classes of
% optimization opportunities.
\subsubsection{Composable Subexpression Support} \label{sec:mqo}
}
\moved{
%  via Multi-query Optimization
As a result of opportunistic evaluation,
there are often many \stmts that
are not completely \rebuttal{executed when issued
by the user}, and are instead executed
in the background asynchronously \rebuttal{during user think time}.
Moreover, by prioritizing
the return of a \rebuttal{prefix or suffix of the results (Section~\ref{sec:prefix})},
often, many \stmts are not computed entirely,
with the computation either deferred (in lazy or eager evaluation),
or being scheduled in the background (in opportunistic evaluation).
Thus, there are many statements that may be
scheduled for execution at the same time.
These statements may operate over similar
or identical subsets of data.
\rebuttal{These overlapping queries that 
can be batch processed make dataframes particularly amenable to}
multi-query optimization (MQO),
e.g.,~\cite{sellis1988multiple,harizopoulos2005qpipe,giannikis2012shareddb,roy2000efficient}. 
\techreport{In fact, some have argued that MQO has
limited applicability in a general relational context:
``One problem of MQO is its
limited applicability (...). In many workloads (...)
there aren't many opportunities
to factor out common subexpressions''~\cite{giannikis2012shareddb},
\rebuttal{and}
``the synchronization of the execution
of queries with common subexpressions
when queries are submitted at different moments in
time''~\cite{giannikis2012shareddb}.
In the dataframe setting, both these reasons for limited
applicability do not hold:
there are often many statements executed essentially
in sync, and there are lots of opportunities
to factor out common subexpressions since
these statements essentially build on top of each other.}
\rebuttal{However, new challenges emerge because
of the new space of operators,
as well as the prioritization of the return
of prefixes/suffixes over the entire result
when requested by the user.
% For example, we may want to share the computation
% of a \code{TRANSPOSE}, since it is particularly
% expensive. 
% Since results are to be produced in an ordered fashion,
% sharing scans as much as possible...
}
% \agp{This needs a "however.." those techniques don't apply 
% directly because? And also we can tone down previous paragraph
% a bit to emphasize the new stuff. }
% \rebuttal{Notice that dataframe query batch processing is only enabled by opportunistic evaluation, which is itself a novel research direction. 
% The interplay between MQO and opportunistic evaluation is fertile ground for novel execution models and optimization techniques.}

}

\moved{
One \rebuttal{simple approach}
is to allow operations
that share inputs to share
scans\techreport{, thereby
reducing the overhead required to
access data}.
We can go even further if
we recognize that many \stmts
are essentially portions of a query
composed incrementally
(e.g.,
\rebuttal{a \code{TRANSPOSE} followed by a \code{PROJECT}
to simulate \code{SELECT}}).
% a cross-product followed by a filter,
% in two separate statements,
% to simulate a join
Therefore, we simply need to construct
a query plan wherein
sub-plans that correspond to
intermediate dataframe results
are materialized as a by-product.
\rebuttal{These intermediates are also  
likely to be reused by the user in the future.}
This poses an interesting conundrum,
because ensuring that the sub-plan results
are materialized ``along the way''
may result in suboptimal
overall plan selection,
which is problematic
when the user cares more about the final
dataframe than intermediates. 
For example,
\rebuttal{
the optimal way to compute a \code{SELECT}
may not be to first compute a \code{TRANSPOSE}
and then do a \code{PROJECT}, even though
this may}
% the overall
% join may not be to first
% compute the cross-product
% and then do the filter, 
% while
% the latter does 
have the benefit
of producing the appropriate intermediate
results.
\rebuttal{Unlike MQO in relational databases,
wherein it is important to share
join subexpressions, here, an even more expensive
operation is \code{TRANSPOSE}---necessitating sharing
if at all possible.}
By using partial results to help
users avoid debugging mistakes,
we may be able to reduce
the importance of constructing
many of the intermediate results
in entirety, unless requested
by the user explicitly.
Moreover, by observing the user's likelihood
of inspecting the intermediates
over the course of many sessions,
we can do a weighted joint optimization
of all query subexpressions,
where the weights for
each intermediate dataframe corresponds to
its importance.
% \agp{Can we talk about other non-relational ops
% here? transposes? For example, an operation on
% a twice transposed relation can be applied in parallel
% with the transpose. If we maintain multiple orientations
% we can use the appropriate orientation for the appropriate
% operation}
}
\moved{
% \devin{One paragraph tech-reported here}
\techreport{
Going one step further, we can try
to jointly optimize not just the evaluation
of intermediate and final result dataframes,
but also the partial or approximate results---a
challenging endeavor.
We can estimate
probabilities for what the user
might do next, e.g., inspect an intermediate,
or compose the next statement,
and the time they may take to do so.
We can couple that with quantifying the benefit
of the user seeing a certain portion
of an intermediate result at a certain time,
to construct a globally optimal query plan.
}
}

\moved{
% \stitle{Materialization and Reuse.}\label{sec:materialization-and-reuse}
\rebuttal{
\subsubsection{Debugging \& Building Queries Incrementally}\label{sec:materialization-and-reuse}
}
The incremental and exploratory nature of dataframe
query construction over the course of a session
leads to nonlinear code paths
wherein the users revisit the same intermediate
results repeatedly as a step towards constructing
just the right queries they want.
In such cases, \rebuttal{intelligently materializing 
key intermediate results can save significant redundant 
computation and speed up query processing.
The optimizer needs to handle the 
}
% the intermediate results
% gain special prominence,
% as the system can try to predict
% which intermediates are likely to be used frequently
% and use such predictions to guide automatic materialization decisions.
% For example, if the user intends to try several
% filters along the way towards constructing
% their intended join, constructing a cross-product intermediate
% first may be sensible.
%The 
trade off between materialization overhead
and the reduced execution time facilitated by availability
of such intermediates % is immediately apparent.
% To use materialized intermediates effectively, we must both {\em (i)
% {\em identify} materialization opportunities} and
% {\em (ii) {\em decide} which such opportunities to take}.
% While the first challenge can be accomplished using standard program analysis,
% the second is less trivial. 
%The dataframe query optimizer needs 
to utilize storage in a way that maximizes
saved compute---small
\rebuttal{intermediate dataframes} that are time-consuming to compute and reused frequently
should be prioritized over \rebuttal{large intermediate dataframes} 
that are fast to compute.
\rebuttal{Note however that materialization
doesn't necessarily need to happen on-the-fly, and can
be also performed in the background asynchronously during
during user think-time.}
\rebuttal{Determining what to materialize 
requires us
%the system can try 
to predict
which intermediates are likely to be used frequently.
% For example, if the user intends to try several
% filters along the way towards constructing
% their intended join, constructing a cross-product intermediate
% first may be sensible.
The prediction algorithm should take into consideration several factors, 
including user intent, past workflows, and operator lineage.

Depending on the underlying intent, 
users can interact with dataframes 
in very different ways. 
A user who is performing data cleaning is 
likely to issue point queries and focus on regions 
with missing or anomalous values; users exploring the 
data for building machine learning models tend to focus on manipulating
columns with high mutual information with the target column,
or more broadly on feature engineering. 
Taking advantage of user intent can lead to highly effective 
materialization and reuse strategies befitting specific access patterns, 
such as in machine learning workflows~\cite{xin2018helix}.
The interactive sessions in dataframe development 
make it possible for the system to infer and adapt to user intent.

User intent inference involves extensive offline analysis 
of workloads with known intents 
as well as online processing of relevant telemetry:
recent Jupyter notebook corpora can provide a 
promising starting point~\cite{ucsd-dataframe-study}.
% In addition to user intent inference, past workloads can also be used to extract access patterns associated with specific intents or users. 
One challenge is that 
unlike SQL workloads, dataframe queries tend to be interleaved 
with non-dataframe operators in the same session, 
which requires special considerations to identify 
the dataframe portion of the workload and to handle 
the interaction between the dataframe system and other frameworks.

Finally, 
dataframe queries in a session often 
build upon one another. 
In the dataflow graph of dataframe queries, 
we are likely to see intermediate results 
that lie on the path to many leaf nodes. 
A simple heuristic is to persist intermediate 
results with high fan-outs; 
more advanced graph analysis techniques 
can be applied to determine prominent intermediate results. 
Opportunistic evaluation can significantly complicate 
the analysis as the execution order 
can differ drastically from the query order.

In terms of costing operators for materialization and reuse, 
the dataframe setting introduces two novel challenges.
Partial views to support fast inspection in conjunction 
with opportunistic evaluation can break up operators 
into multiple partial operators evaluated at different times, 
motivating the need for short and long term costs 
on partial views for each operator. 
The materialization and reuse decisions 
derived from these costs can feed back into the decisions 
on filtering for partial views or delaying evaluation. 
For example, if several queries based on a new sort order 
require immediate feedback in the near future, 
it might be prudent to incur a delay on the first query 
to materialize the new sort order in its entirety, 
in order to significantly speed up subsequent queries on the new order through reuse. 
Of course, being able to make such decisions 
hinges on the ability to predict future reuse as discussed above.
Secondly, the constantly growing dataflow graph 
requires eviction of old materialized results from memory. 
The interesting challenge in the dataframe context 
is that future reuse is determined by both what 
the user will do in the future and what the opportunistic evaluator 
will choose to compute, with the former being purely speculative 
and the latter being known within the system. 
We can reconcile the ``two futures'' by passing 
the model we build of the future workflow 
to the opportunistic scheduler 
for unified materialization/reuse planning.

% Finally, given the equivalence of rows and columns in the dataframe data model and the frequent invocation of \code{TRANSPOSE}, materializing the same dataframe in both row and column major layouts can help speed up future queries by avoiding physical transposes. The optimizer will be able to freely choose either layout for an operator instead of having to shoehorn a suboptimal plan on the inferior layout to avoid physical transpose. The caveat is that the materialization for the two orientations could potentially become out of sync. The optimizer needs to carefully weigh the trade-off between the benefit of avoiding physical transposes and overhead of extra storage and syncing the two materializations.

}

% Although previous work has shown that the optimal reuse policy given
% a dataflow graph can be identified in PTIME~\cite{xin2018helix},
% %Storage can refer to either memory or disk,
% %depending on whether reuse takes place within a single session
% %or across multiple sessions.
% a special wrinkle in the dataframe setting is that we may be able
% to get away with partial view materialization,
% e.g., materializing a prefix of the cross-product,
% to be able to efficiently support prefix computation
% of the results of several filters applied
% to that cross-product.
% \agp{Reword this to say that.. OK I forgot my train of thought here. will try again later}
}
\moved{

% \dorx{This paragraph is standard DB pipelining and mat. Candidate for removal.}
% \devin{agree candidate for TR}
\techreport{
Another approach to speed up dataframe queries
would be to defer the creation of new 
dataframes as a result of queries
and instead allow for the results
of dataframe queries to be essentially
non-materialized ``views''.
This could be useful, for example,
when a dataframe query essentially adds
a new derived column for feature engineering.
In this case, we don't actually add the derived
column and create a new dataframe,
simply recording the operations instead,
and materializing the result on-demand.
Deferring the operations also opens up
opportunities for pipelining through subsequent
operations, saving overall computation costs.
In fact, the Vaex project~\cite{vaex},
which is a dataframe-like system (as we described in Section~\ref{sec:dataframedef})
that supports querying on HDF5 files,
implements virtual columns.
With virtual columns,
the column is not actually materialized
until required for output, printing, or for a query.
In cases where the computation
that creates a column is expensive,
virtual columns will
need to be paired with intelligent caching mechanisms
that prioritize caching
columns that were expensive to generate.
}
}
\balance
\section{Related work}\label{sec:ext}
While our focus on \pandas is driven 
by its popularity, in this section, 
we discuss other existing dataframe 
and dataframe-like implementations. 
Table~\ref{tab:comparison} outlines the features
of these dataframe and dataframe-like implementations.
We will discuss how existing dataframe implementations fit into our framework, 
thus showing how our proposed research is applicable to these systems.

%New outline here: \url{https://docs.google.com/document/d/1wK0O31EsR4LzH6PIaaeIDPVh0u8wEDkPL0YGawl2SC0/edit?usp=sharing}

% Designed specifically to support statistical modelling, dataframes were first introduced in the S language~\cite{chambers1992statistical}, a predecessor to the R language. 
% %Both the S and Scheme language heavily influenced the development of the R language, which popularized the dataframe at the center of our discussion~\cite{ihaka1996r}. 
% In S and Scheme, the \textit{frame} in dataframe refers to the notion of \textit{environment frames}. 
% An \textit{environment} is the association between symbols and values, and an \textit{environment frame} is a list of symbol-value pairs.\dlee{unclear what symbol mean here, can we just say that these are key-value pairs?} 
% The term dataframe thus captures the dual roles of being a data matrix and a list of columns and rows variables.
% While bearing semblance to relations, dataframes have evolved independently of RDBMS, thus motivating the need for unification.
\rebuttal{\stitle{Data model and algebra.}}
% \todo{Add discussion here about related works: array databases, other databases, etc.}
To the best of our knowledge,
an algebra for dataframes has never
been defined previously.
Recent work by Hutchinson et al.~\cite{lara, laradb}
proposes an algebra called Lara that combines
linear and relational algebra, exposing 
only three operators:
\code{JOIN}, \code{UNION}, and \code{Ext} (also known as ``flatmap'');
however, \removed{the} \rebuttal{dataframe} metadata manipulation operators \removed{below}
\rebuttal{are not supported.}
\removed{would not be possible in Lara without
placing the metadata as part of the data.}
Other differences stem from the flexible data model
and lazily induced schema.
\removed{That said,} \rebuttal{We will draw on
Lara as we continue to refine our algebra.}

% \rebuttal{
% Array databases are conceptually similar to dataframes in the
% context of the order of the data, where arrays have an implicit
% order. However, many of the 
% }

\rebuttal{\stitle{Dataframe Implementations: R.}} %: The R Dataframe}
As we discussed in Section~\ref{sec:dataframe-history},
the R language (and the S language before it),
both support dataframes in a manner similar to \pandas
\rebuttal{and can be credited for initially popularizing
the use of dataframes for data analysis}~\cite{ihaka1996r}. 
\rebuttal{R is still quite popular, especially
among the statistics community.}
An R dataframe is a 
list of variables, 
each represented as a column, with the same number of rows. 
While both the rows and columns in an R dataframe have names, 
row names have to be unique;
thus the \pandas dataframe is more permissive than the R one.
%The column names correspond to the column labels described in Section~\ref{sec:definitions}, while the uniqueness of row names can be achieved with the row metadata\agp{?}. 
As shown in Table~\ref{tab:comparison}, R supports all of the operations in our algebra. 
The R dataframe fully captures our definition of a dataframe, and 
thus, implementational support of R dataframes requires only conforming the R API 
to our proposed algebra.
% \todo{Simon: show an example of how to do this.}
% \simon{? I don't think we need to add anything here. }
% \dlee{There shouldn't be a checkmark in the R column for Eager execution? Can somebody else verify this and change Table 2?}
External R packages such as readr, dplyr, and ggplot2 operate on R dataframes and provide functionalities such as data loading, transformation, and visualization, similar to ones from the \pandas API~\cite{tidyverse-R,tidy-data}.

\stitle{Dataframe-like Implementations.} %: Spark, Koalas, Dask, and ORMs}
Some libraries provide a functional or 
object-oriented programming layer 
on top of relational algebra. 
These libraries include SparkSQL dataframes~\cite{sparksql-func}, 
SQL generator libraries like QueryDSL~\cite{querydsl} and JOOQ~\cite{jooq}, 
and object relational-mapping systems (ORMs) such as Ruby on Rails~\cite{rubyrails} 
and SQLAlchemy~\cite{sqlalchemy}. 
All of these systems share some of the benefits with respect to 
incremental query construction mentioned in Section~\ref{sec:interactivity}. 
However, they generally do not support the richness and expressiveness 
of \removed{the} dataframes \removed{data model and algebra}, 
including \removed{keys aspects like} ordering of rows, symmetry between rows and columns,
and operations such as transpose.

SparkSQL and Dask are scalable dataframe-like systems
that take advantage of distributed computing to handle large datasets. 
However, as shown in Table~\ref{tab:comparison}, 
Spark and Dask do so at the cost of limiting 
the supported dataframe functionalities. 
For example, a dataframe in SparkSQL does not treat columns
and rows equivalently and requires a predefined schema.
As a consequence, SparkSQL does not support \code{TRANSPOSE}
and is not well optimized for dataframes 
where columns substantially outnumber rows.
Thus, SparkSQL \removed{does not match our definition 
of a dataframe and is much} is closer to a relation 
than a dataframe.
Koalas~\cite{koalas}, a wrapper on top of the SparkSQL API, 
attempts to be more dataframe-like in the API 
but suffers from the same limitations\removed{as the underlying SparkSQL framework}.
\techreport{
\par \removed{On the other hand,} Dask enables distributed 
processing by partitioning along the rows 
and treating each partition as a separate ``dataframe'', 
thus acting as a ``meta-dataframe''.
Since ordering and transpose are ill-defined for a group of dataframes, 
Dask fundamentally cannot support \removed{such} operations that rely on row-ordering. 
The set of operations supported \removed{in Dask 
is determined by which dataframe operations}
\rebuttal{are restricted to those that}
can be combined into a single output based on the resulting, 
constituent dataframes. 
These include \rebuttal{embarassingly parallel} operations, such as filter, aggregation, groupby, \rebuttal{and} join.
}

\techreport{
Unlike these systems, \modin treats the dataframe data model and algebra
as first-class citizens, as opposed to a means to enable distributed processing,
addressing challenges in dataframe processing
in systems like \pandas and R at scale,
while not sacrificing the convenient functionalities
that have made dataframes so popular. We advocate that our research vision around the data model proposed in this paper is a key component towards this more holistic approach for optimizing dataframe systems.
}

\begin{table}[t]
\centering
\small
\begin{tabular}{|l|l|l|l|l|l|}
\hline
{\bf Feature}         & \cellcolor{blue!25}{\bf Modin}      & \cellcolor{blue!25}{\bf \Pandas}    & \cellcolor{blue!25}{\bf R}           & \cellcolor{red!25}{\bf Spark}      & \cellcolor{red!25}{\bf Dask}        \\ \hline
Ordered model   & \checkmark & \checkmark & \checkmark  & \checkmark$^\dagger$ &             \\ \hline
Eager execution & \checkmark & \checkmark & \checkmark$^+$  &            &             \\ \hline
Row/Col Equivalency & \checkmark & \checkmark & \checkmark  &            &             \\ \hline
Lazy Schema  & \checkmark & \checkmark & \checkmark  &            & \checkmark    \\ \hline
Relational Operators   & \checkmark & \checkmark & \checkmark  & \checkmark & \checkmark  \\ \hline
MAP             & \checkmark & \checkmark & \checkmark  & \checkmark & \checkmark  \\ \hline
WINDOW          & \checkmark & \checkmark & \checkmark  & \checkmark & \checkmark  \\ \hline
TRANSPOSE       & \checkmark & \checkmark & \checkmark  &            &             \\ \hline
TOLABELS       & \checkmark & \checkmark & \checkmark  &            & \checkmark* \\ \hline
FROMLABELS    & \checkmark & \checkmark & \checkmark  &            &             \\ \hline
\end{tabular}
\vspace{-5pt}
\caption{Table of comparison between dataframe and dataframe-like implementations. Blue indicates
dataframe systems, red indicates dataframe-like implementations.
$\dagger$: Spark can be treated as ordered for some operations.
$+$: R dataframe operators can be invoked lazily or eagerly.
*: Dask sorts the dataframe by the row labels after \texttt{TOLABELS}.
}\label{tab:comparison}
\vspace{-15pt}
\end{table}

% Section 10
%!TEX root = modin-vision.tex

\section{Conclusion}\label{sec:conclusion}

In recent years, the convenience of dataframes
have made them 
the tool of choice 
for data scientists 
to perform a variety of tasks, from data loading, cleaning, 
and wrangling to statistical modeling and visualization.
Yet existing dataframe 
systems like \pandas 
have considerable difficulty in providing interactive responses
on even moderately-large datasets of less than a gigabyte.
\rebuttal{
This paper outlines our research agenda
for making dataframes scalable,
without changing the functionality or usability
that has made them so popular. 
Many fundamental assumptions made by 
relational algebra are entirely discarded in favor of new ones for dataframes,
including rigid schemas, an unordered data model, rows and columns being
distinct, and a compact set of operators. 
Informed by our experience in developing \modin,
a drop-in replacement for \pandas,
we described a number of research challenges
that stem from revisiting familiar data management problems,
such as metadata management, layout and indexing, and
query planning and optimization, under these new assumptions.
As part of this work, we also proposed a
candidate formalism for dataframes,
including a data model 
as well as a compact set of operators,
that allowed us to ground our research directions
on a firm foundation.
}
% This paper outlines several directions
% towards improving dataframe systems by making
% them more scalable. 
% Our central goal is to {\em adapt
% traditional optimization 
% techniques pioneered by the database community
% to dataframes without
% sacrificing functionality or usability}.
% \rebuttal{
% To achieve this goal,
% we developed \modin,
% which attempts to bring database optimizations to dataframes.
% However} 
% optimizing dataframes is challenging 
% due to its imperative, incrementally composed querying modality,
% as well as a large and redundant set of query primitives.
% To this end, we proposed a formal definition 
% and candidate data model for dataframes,
% and contributed a small set of operators 
% that can be used to express the \pandas API.
% \removed{
% We introduced our ongoing effort 
% to bring database optimizations
% to dataframes, \modin.
% }
% We outlined a number of new research directions to make
% dataframes more efficient,
% targeting optimizations that revisit every step of 
% a typical stack of a database-like system,
% as well as those that stem from how dataframes are leveraged---in an incrementally
% constructed manner, with inspection of intermediate results.
\rebuttal{We hope our work serves as a roadmap and a call-to-action 
for others in the database
community to contribute to this emergent, exciting, and challenging
research area of 
scalable dataframe systems
development.}
% start contributing to the 
% important area of developing better dataframe systems
% for exploratory data analysis.

\techreport{
\subsection*{Acknowledgments}
We would like to acknowledge those who have made significant contributions to the \modin codebase: we thank Omkar Salpekar,
Eran Avidan, Kunal Gosar, GitHub user ipacheco-uy, Alex Wu, and Rehan Sohail Durrani.
We also thank Ion Stoica for 
initial discussions and encouragement.
}
%}

\papertext{\newpage}

\bibliographystyle{abbrv} % vldb
\bibliography{modin-vision.bib}

\end{document}